# Reinforcement Learning Based User-Guided Motion Planning for Human-Robot Collaboration

Tian Yu[a], Qing Chang[a,*]

[a]*Department of Mechanical and Aerospace Engineering, University of Virginia, Charlottesville, VA, 22904, US*

**Abstract**

Robots are good at performing repetitive tasks in modern manufacturing industries. However, robot motions are mostly planned and preprogramed beforehand with a notable lack of adaptivity to task changes. Even for slightly changed tasks, the whole system must be reprogrammed by robotics experts. Therefore, it is highly desirable to have a flexible motion planning method, with which robots can adapt to certain task changes in unstructured environments, such as production systems or warehouses, with little or no intervention needed from non-expert personnel. In this paper, we propose a reinforcement learning (RL) based motion planning method to enable robots to automatically generate their motion plans across different tasks by learning from a few kinesthetic human demonstrations. To achieve adaptive motion plans for a specific application environment, e.g., desk assembly or warehouse loading/unloading, a library is built by abstracting features of common human demonstrated tasks. The definition of semantical similarity between features in the library and features of the new task is proposed and further used to construct the reward function in RL. A Q-learning algorithm is applied to the motion planning policy training. The RL policy would automatically generate motion plans for a new task if it determines that new task constraints can be satisfied with current library, and otherwise request additional human demonstrations. Multiple experiments conducted on common tasks and scenarios demonstrate that the proposed RL-based user-guided motion planning method is effective.

## 1. Introduction

With the development of Industry 4.0, there is an increasing demand for robots to work adaptively and smartly with humans in industrial settings (El Zaatari et al., 2019). However, currently, the motions of industrial robots are still largely preprogrammed to perform certain repetitive tasks. When tasks sightly change, robots need to be reprogrammed, which often requires considerable robotic expertise. This would significantly impair the efficiency of industrial robots that constantly encounter new tasks, and hence limit

Tian Yu
Email address: ty2yy@vriginia.edu

*Corresponding Author: Qing Chang
Email address: qc9nq@virginia.edu

their use in quite a few industrial scenarios, e.g., small-batch manufacturing that processes highly customized products (Jurczyk-Bunkowska, 2020; Poon et al., 2011). Therefore, how to enable robots to collaborate with non-expert personnel and automatically plan adaptive motions for different tasks is a nontrivial and challenging research problem in today's robotics research (Aleotti & Caselli, 2006; Kragic et al., 2018; Laha, Figueredo, et al., 2021). In this paper, our goal is to develop a scalable and adaptive motion planning method to automatically generate motion plans for new robotic manipulation tasks without manually reprogramming robots. To achieve this goal, we propose to represent a task by a sequence of critical constraints, and combine human demonstrations with motion planning algorithms to generate motion plans to fulfill those constraints.

In literatures, the use of human demonstration to teach a robot is often referred to as Learning from Demonstration (LfD). An important question in LfD is how to acquire demonstrations for learning, e.g., using vision-based sensors, kinesthetic, and data gloves or master-slave systems (Argall et al., 2009; Zhu & Hu, 2018). In this paper, we opt to use kinesthetic demonstrations for the following considerations. On one hand, for kinesthetic demonstrations, there is no correspondence issue between the kinematic structure of the demonstrating system and the follower robot. On the other hand, learning from kinesthetic demonstrations can potentially benefit from a large variety of existing approaches in learning motion from data, which can be classified as follows: (a) demonstrated trajectory decomposition (Hwang et al., 2003; Madridano et al., 2021), (b) nonlinear regression techniques (Aleotti & Caselli, 2006; Calinon et al., 2007; Kulić et al., 2008), and (c) dynamical systems based approach (Gribovskaya et al., 2011; Ijspeert et al., 2013; Jokić et al., 2022)

Specifically, the trajectory decomposition approaches (Hwang et al., 2003; Madridano et al., 2021) use spline functions to decompose the trajectories. These methods ignore the noises in the demonstration, which may be nontrivial especially when the motion information is obtained through vision or teleoperation. Nonlinear regression techniques use statistical techniques to incorporate the uncertainty of sensing in the estimation. For example, (Aleotti & Caselli, 2006) uses Hidden Markov Model for trajectory selection and Non-Uniform Rational B-Splines (NURBS) for trajectory approximation. A data-driven approach using Gaussian Mixture Model (GMM) is adopted in (Calinon et al., 2007). However, these statistical approaches require multiple demonstrations. Furthermore, neither the statistical approaches nor the spline decomposition takes the kinetic transformations of the underlying task space, i.e., $SE(3)$, into account. The dynamical systems-based approach, or dynamical motion primitives (DMPs), on the other hand, can learn from single examples (Ijspeert et al., 2013). However, in these settings, most works assume that there is a dynamical system modeling each degree-of-freedom (DoF) of the end effector. In (Gribovskaya et al., 2011), the authors assume a unit quaternion representation of the rotation space. However, they assume decoupled primitives for the four parameters of the unit quaternion and then perform a normalization. The generalization of DMPs is predicated on the region of attraction of the dynamical system used. In the case of orientations, there is no clear characterization of the region of attraction of the dynamical system on the group of rigid body rotations, i.e., $SO(3)$. Therefore, the generalization capability of these methods is not clear when both the position and orientation of the end effector are relevant for the task. Therefore, it is imperative to develop a systematic LfD method to ensure that the underlying task space structure of $SE(3)$ is conformed and exploited during motion generation.

In general, techniques for motion planning and robot control can be divided into Joint-space based approaches and Task-space based approaches (Kavraki et al., 1996; Khatib, 1987). Joint space-based motion planning approaches handle the planning problem and compute the motion directly in the joint space of the robot. The strength of joint space methods lies in finding feasible paths that avoid obstacles (Berenson et al., 2009). However, handling task constraints in joint-space based planning approaches is quite complicated (Jaillet & Porta, 2012) because they lead to nonlinear constraints in joint angles.Task space based planning approach is historically older than joint-space based approaches and rose out of the resolved motion rate control (RMRC) in (Klein & Huang, 1983). Related to the task space based planning approaches are the operational space based control approaches (Nakanishi et al., 2008), where the redundancy resolution may be done at either the velocity level or acceleration level.

Recently, (Laha, Rao, et al., 2021) develop a user-guided motion planning method that learns from only one human demonstration to generate motion plans for semantically similar task instances. They first compute an "imitated path" in the task space by replicating the human demonstration based on the goal position of the new task instance. Then, use the "imitated path" as the guidance of Screw Linear Interpolation (ScLERP) until the current configuration of the end-effector finally blends into the imitated path. However, task constraints before the current configuration blends into the imitated path are not guaranteed to be satisfied. In addition, the only explicit task constraints considered in this work is the goal position of the task instance. Other explicit task constraints like positions and orientations of some critical configurations and environment conditions are not considered. More importantly, the method in (Laha, Rao, et al., 2021) is only able to generate motion plans when the whole new task is semantically similar to the demonstrated one, e.g., moving the same water bottle, but to a different goal position. As a matter of fact, if we decompose a robot motion into several portions appropriately, we might find that some of those portions could be semantically similar across different tasks. For example, a portion of robot motions in a transferring task could be similar to that in a stacking task. Based on this observation, it is possible to generate motion plans for completely new tasks purely based on old tasks that the robot has been taught without additional demonstrations or programming efforts from human.

In this paper, we develop a motion planning method that can enable the robot to learn from one or even multiple human demonstrations to generate adaptive motion plans for new manipulation tasks in a certain manufacturing environment. First, we define a syntax to specify the manipulation task in an assembly and loading/unloading scenario considering both explicit task constraints, e.g., critical configurations of the end-effector, and environment constraints, e.g., dimension and location of the obstacle. Next, we build a library to store human demonstrated features which are embedded in screw transformation throughout demonstrations. The same method can also be applied to abstract features of the manipulation task. A criterion to identify semantical similarity of a human demonstration and certain part of the manipulation task is defined. Based on this criterion, the appropriate features of human demonstrations will be mapped to the manipulation task. Therefore, to generate a motion plan for the manipulator to satisfy both explicit and implicit task constraints is equivalent to mapping appropriate features in the library to corresponding parts of the manipulation task. In this work, we formulate the motion planning problem in $SE(3)$ into a Markov Decision Process (MDP) framework and use the Q-learning method to train a general motion planning policy to generate adaptive motion plans in $SE(3)$ for different tasks in the same assembly and loading/unloading environment. Finally, inverse kinematics (IK) is used to calculate corresponding motion plans in the joint space to control the robot to execute learned motion plans.

Thus, the main contributions of this paper are: (1) Propose a scheme to specify manipulation tasks based on explicit task requirements and environment constraints; (2) Propose a method to abstract features from human demonstrations through kinematic transformation; (3) Set up a criterion to identify semantical similarity between features of human demonstrations and manipulation tasks; (4) Develop a mapping method to scale features of human demonstrations to semantically similar task segments; (5) Formulate the learning from demonstration and motion planning problem in an MDP framework, and implement a Q-learning algorithm to solve the formulated problem effectively.

The reminder of this paper is organized as follows: The mathematical preliminaries are introduced in Section 2. The learning from demonstration and motion planning problem is stated in Section 3. In Section 4, the problem is formulated as an MDP and solved by the Q-learning algorithm. Section 5 describes our approach of inverse kinematics. Case studies and conclusions are provided in Section 6 and 7 respectively.

## 2. Mathematical Background

In this paper, the joint space or configuration space is represented by $\mathcal{J}$, which is the set of all joint angles of the robot manipulator. $SE(3)$ denotes the Special Euclidean group of 3, which represents the task space contains all rigid body motions (i.e., rotations and translations) (Selig, 2005).

To describe the rigid body rotation and translation in $SE(3)$, a dual quaternion representation is adopted in this paper. A quaternion is defined as (Figueredo, 2016):

$$q = a + b\hat{\boldsymbol{\imath}} + c\hat{\boldsymbol{\jmath}} + d\hat{\boldsymbol{k}} \tag{1}$$

where $a, b, c$ and $d$ are scalars and $\hat{\boldsymbol{\imath}}, \hat{\boldsymbol{\jmath}}, \hat{\boldsymbol{k}}$ are orthogonal unit vectors. The conjugate of quaternion $q$ is defined as:

$$q^* = a - b\hat{\boldsymbol{\imath}} - c\hat{\boldsymbol{\jmath}} - d\hat{\boldsymbol{k}} \tag{2}$$

Let $q_1 = a_1 + b_1\hat{\boldsymbol{\imath}} + c_1\hat{\boldsymbol{\jmath}} + d_1\hat{\boldsymbol{k}}$ and $q_2 = a_2 + b_2\hat{\boldsymbol{\imath}} + c_2\hat{\boldsymbol{\jmath}} + d_2\hat{\boldsymbol{k}}$ be two quaternions, their product, called the Hamilton product, is given as:

$$\begin{aligned} q_1 \otimes q_2 = {} & a_1a_2 - b_1b_2 - c_1c_2 - d_1d_2 + (a_1b_2 + b_1a_2 + c_1d_2 - d_1c_2)\hat{\boldsymbol{\imath}} + (a_1c_2 - b_1d_2 + c_1a_2 + d_1b_2)\hat{\boldsymbol{\jmath}} \\ & + (a_1d_2 + b_1c_2 - c_1b_2 + d_1a_2)\hat{\boldsymbol{k}} \end{aligned} \tag{3}$$

The norm of $q_1$ is defined as:

$$\|q_1\| = \sqrt{q_1q_1^*} = \sqrt{a_1^2 + b_1^2 + c_1^2 + d_1^2} \tag{4}$$

The Euclidean distance between two quaternion is defined as (Huynh, 2009):

$$\phi(q_1, q_2) = \min\{\|q_1 - q_2\|, \|q_1 + q_2\|\} \tag{5}$$

Unit quaternions are a singularity-free representation of $SO(3)$ and are used to represent rotations. The relationship between Euler angles is shown in Fig. 1, and the unit quaternion $q_{unit}$ is represented as:

$$q_{unit} = \begin{aligned} &\cos\left(\frac{\varphi}{2}\right)\cos\left(\frac{\theta}{2}\right)\cos\left(\frac{\psi}{2}\right) + \sin\left(\frac{\varphi}{2}\right)\sin\left(\frac{\theta}{2}\right)\sin\left(\frac{\psi}{2}\right) + \\ &\left(\sin\left(\frac{\varphi}{2}\right)\cos\left(\frac{\theta}{2}\right)\cos\left(\frac{\psi}{2}\right) - \cos\left(\frac{\varphi}{2}\right)\sin\left(\frac{\theta}{2}\right)\sin\left(\frac{\psi}{2}\right)\right)\hat{\boldsymbol{\imath}} + \\ &\left(\cos\left(\frac{\varphi}{2}\right)\sin\left(\frac{\theta}{2}\right)\cos\left(\frac{\psi}{2}\right) + \sin\left(\frac{\varphi}{2}\right)\cos\left(\frac{\theta}{2}\right)\sin\left(\frac{\psi}{2}\right)\right)\hat{\boldsymbol{\jmath}} + \\ &\left(\cos\left(\frac{\varphi}{2}\right)\cos\left(\frac{\theta}{2}\right)\sin\left(\frac{\psi}{2}\right) - \sin\left(\frac{\varphi}{2}\right)\sin\left(\frac{\theta}{2}\right)\cos\left(\frac{\psi}{2}\right)\right)\hat{\boldsymbol{k}} \end{aligned} \tag{6}$$

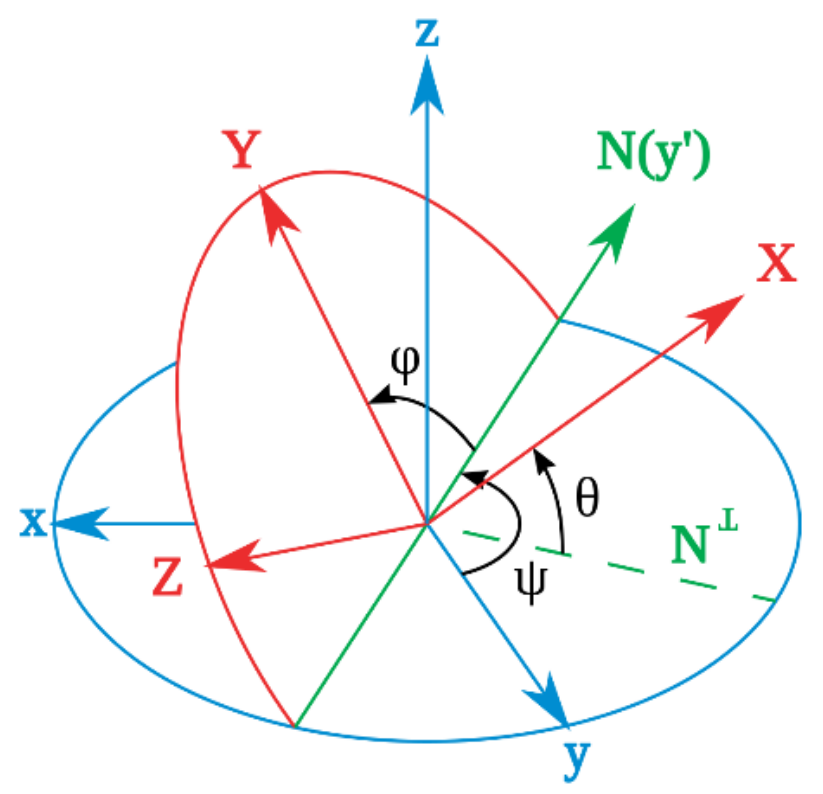

**Figure. 1** Euler angles in “ZYX” sequence. The angle rotation sequence is $\psi, \theta, \varphi$. Yaw-$\psi$: rotation about the Z-axis; Pitch-$\theta$: rotation about the new Y-axis; Roll-$\varphi$: rotation about the new X-axis.

Rigid body transformations can be elegantly described by dual quaternions as they encode both rotation and translation (Selig, 2005). A dual quaternion $\boldsymbol{D}$ is defined as:

$$\boldsymbol{D} = \boldsymbol{d}_r + \frac{1}{2}\epsilon \boldsymbol{d}_t \otimes \boldsymbol{d}_r \tag{7}$$

where $\epsilon \neq 0$, but $\epsilon^2 = 0$. In this definition, the pure translation of the rigid body is represented by the quaternion $\boldsymbol{d}_t$ (shown in Fig. 2 (a)) which is denoted as:

$$\boldsymbol{d}_t = (0, \hat{\boldsymbol{t}}) \tag{8}$$

where $\hat{\boldsymbol{t}} = t_x\hat{\boldsymbol{\imath}} + t_y\hat{\boldsymbol{\jmath}} + t_z\hat{\boldsymbol{k}}$ is the translation vector in $SE(3)$. The unit quaternion $\boldsymbol{d}_r$ (shown in Fig. 2 (b)) representing the pure rotation of the rigid body can also be expressed as:

$$\boldsymbol{d}_r = \cos\left(\frac{\phi}{2}\right) + \hat{\boldsymbol{n}}\sin\left(\frac{\phi}{2}\right) \tag{9}$$

where $\hat{\boldsymbol{n}} = n_x\hat{\boldsymbol{\imath}} + n_y\hat{\boldsymbol{\jmath}} + n_z\hat{\boldsymbol{k}}$ is a unit vector in $SE(3)$ representing the rotation axis, and $\phi$ is the rotation angle.

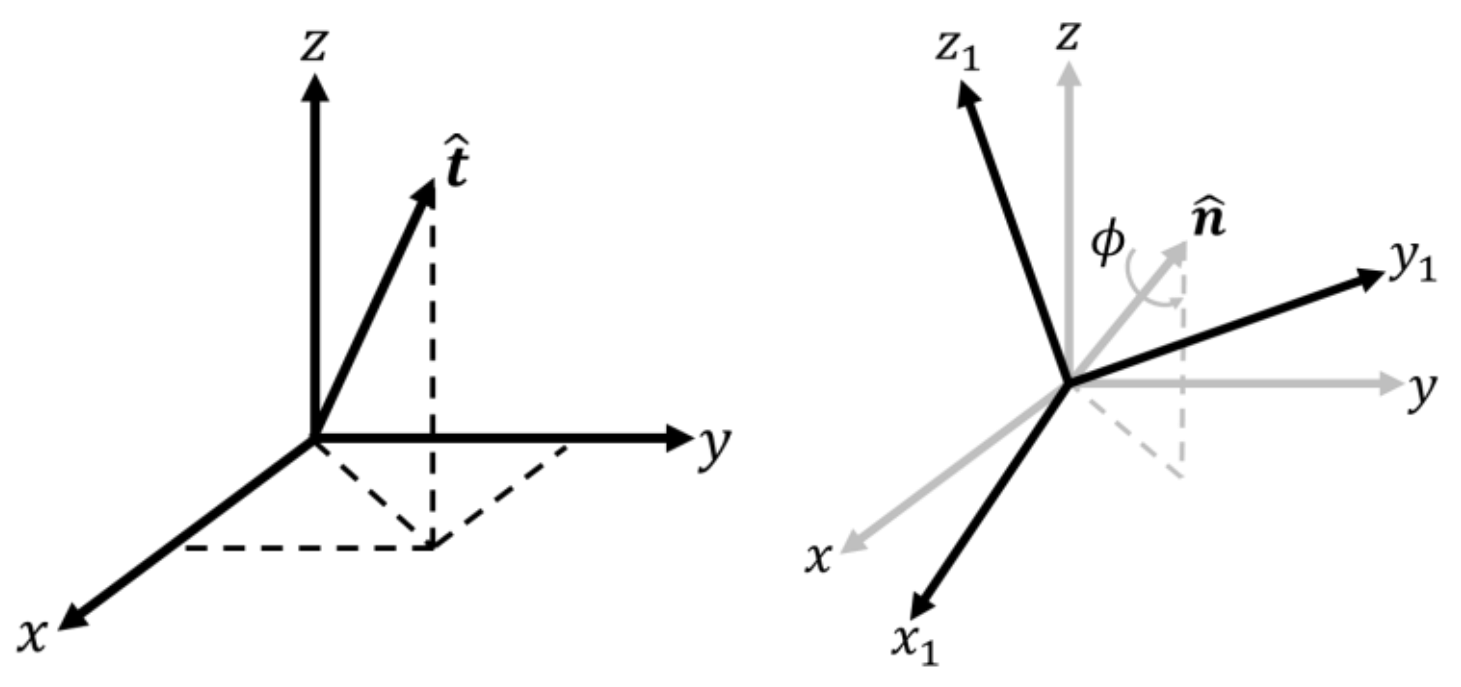

(a) (b)

**Figure. 2** Quaternion representations of rigid body translation and rotation. (a) Translation quaternion $\boldsymbol{d}_t = t_x\hat{\imath} + t_y\hat{\jmath} + t_x\hat{k}$. Note that both the real and imaginary axes form an orthogonal basis, but the real axis is omitted. (b) Rotation angle $\phi$ around the unit rotation axis $\hat{\boldsymbol{n}}$, represented by $\boldsymbol{d}_r = \cos\left(\frac{\phi}{2}\right) + \hat{\boldsymbol{n}}\sin\left(\frac{\phi}{2}\right)$.

Using this $\boldsymbol{d}_r$, any vector $\hat{v}$ can be rotated an angle $\phi$ about the axis $\hat{\boldsymbol{n}}$ by using the quaternion sandwich $\boldsymbol{d}_r\hat{v}\boldsymbol{d}_r^*$, and $\boldsymbol{d}_r^*$ is the conjugate of $\boldsymbol{d}_r$. For more quaternion manipulations, we refer readers to (Figueredo, 2016).

### 3. Problem Formulation

In this paper, a robot is required to do manipulation tasks with explicit and implicit constraints on end-effector configurations during a motion. We assume that the robot has basic capability to move its end-effector from one configuration to another in the absence of any constraints. Our goal is to develop a method to enable the robot to use human demonstrations from a few common tasks stored in a scenario specific library and to plan point-to-point motions for other new tasks. The human demonstration library can be formed for each specific working scenarios, such as desk assembly library, warehouse sorting library etc. On human demonstrations, we take (one-time) kinaesthetic demonstration for one type of tasks and the information is observed by the screw transformation throughout the trajectory. Using such screw transformation, the feature of the manipulation task can be abstracted in $SE(3)$. To select the appropriate human demonstration to be learnt from for the manipulation task, criteria are built by comparing the screw transformation of the human demonstration and corresponding screw transformation of a few critical configurations of the task. Following the criteria, features of appropriate human demonstrations can be mapped to the new manipulation task in $SE(3)$.

*3.1. Specify Manipulation Tasks for the Robot*

First of all, the manipulation task comprehensible to the robot needs to be specified. Some research (Konidaris et al., 2018; Wang et al., 2018) describes a task, e.g., "open the door", using a symbolic vocabulary based on the high-level Planning and Domain Definition Language (PDDL), e.g., {preposition of the door, precondition: door closed, effect: door open}, to build a task library. These high-level commands lack detailed kinematics information and cannot be directly translated to actionable information for the lower-level robot manipulation. (Hauser & Ng-Thow-Hing, 2011) sample reasonable modes for motions of a humanoid robot by specifying the manipulation as the starting and goal configurations, together with detailed transition configurations during the entire task in the task space. However, for a general manipulation task in practice, the specific transition from one configuration to another in both $SE(3)$ and $\mathcal{J}$ is not known to the robot. Therefore, in this paper, the manipulation task is specified as a set of critical positions obtained from the known constraints based on task requirements (e.g., a specific goal position) and the environment constraints (e.g., the location of an obstacle), and the associated orientation tolerance for the end effector represented in $SE(3)$ at these positions.

Starting from an initial configuration of the end-effector, a manipulation task $\boldsymbol{TK}$ is defined as a sequence of $n$ critical configurations:

$$\boldsymbol{TK} = \{con_1, con_2, \ldots, con_n\} \tag{10}$$

where $con_i,\ i = 1, \ldots, n$, is a tuple of two $< \boldsymbol{P}_i, \boldsymbol{\theta}_i >$. In this tuple, $\boldsymbol{P}_i = [x_i, y_i, z_i]^T$ is a vector in $SE(3)$ that specifies the position of $con_i$, $\boldsymbol{\theta}_i = [\theta_{i1}, \theta_{i2}, \theta_{i3}]^T,\ \theta_{l1} \le \theta_{i1} \le \theta_{u1}, \theta_{l2} \le \theta_{i2} \le \theta_{u2}, \theta_{l3} \le \theta_{i3} \le \theta_{u3}$, is a unit vector in $SE(3)$ that defines Euler angles of $con_i$, and $\theta_{l1}, \theta_{l2}, \theta_{l3}, \theta_{u1}, \theta_{u2}, \theta_{u3}$ are the lower and upper bounds for corresponding Euler angles. For this task specification, only a few task-related explicit requirements are given. However, there might be implicit task constraints (e.g., maintain a certain orientation from one configuration to another) need to be complied by the robot. Our goal is to find a motion plan to satisfy all the task requirements.

For example, a task $\boldsymbol{TK}$ of transferring a cup of water shown in Fig. 3 can be specified by four critical configurations, $con_1, con_2, con_3, con_4$, as summarized in Table 1. In this task, the end-effector is required to move a cup of water from the starting configuration $con_1$ to the goal configuration $con_4$ while maintaining the cup upward. It is noted that, to move the cup without spilling water out, the pitch and roll angles of the end-effector need to be constrained within a certain range while the yaw angle does not need to be constrained. Therefore, $\gamma$ in Table 1 can be any number from $-2\pi$ to $2\pi$, and $\theta_{i1}$ and $\theta_{i2}$ for each $\boldsymbol{\theta}_i$ are set to be 0 for simplicity. Furthermore, if the orientation of the end effector is required to remain the same during the entire transferring, then $\gamma$ can be also specified to be 0.

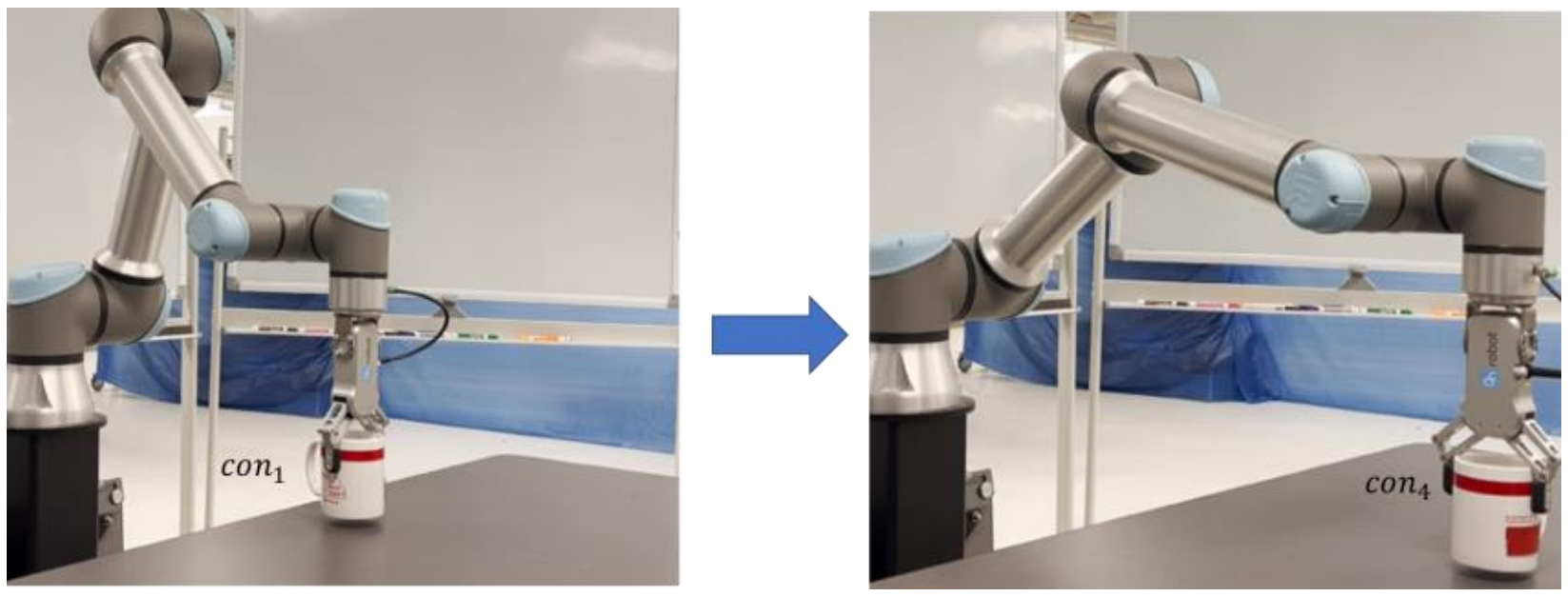

**Figure. 3** Critical configurations of a transferring task.

**Table 1**. Critical Configurations of $\boldsymbol{TK}$

| | $con_1$ | $con_2$ | $con_3$ | $con_4$ |
|---|---|---|---|---|
| $\boldsymbol{P}$ | (-0.2,0,0.6) | (-0.3,0,0.6) | (-0.4,0,0.6) | (-0.5,0,0.6) |
| $\boldsymbol{\theta}$ | (0,0,$\gamma$) | (0,0,$\gamma$) | (0,0,$\gamma$) | (0,0,$\gamma$) |

**Remark 1**: In this paper, we only provide a syntax on how to specify a manipulation task. The practical task specification is quite a heuristic practice. If the user has sufficient knowledge on the requirements of a new task and the corresponding environment constraints, one may better define the critical positions and the associated end effector orientations. This will help the robot better understand the new task and facilitates a more effective learning in performing the new task.

### *3.2. Build the Library of Human Demonstrated Features*

To facilitate the learning from human demonstration, we want to build a library to include primitive or common tasks for specific working scenarios. For example, for a workshop that works on assembling height-adjustable desks, the tasks of twisting screws clockwise and placing screws into assembly holes are common or primitive tasks and can be included in the library for this workshop or working scenario. To demonstrate these primitive tasks, the human operator can physically hold and move the robot's end-effector from the initial configuration to the goal configuration. Throughout the transformation trajectory, the explicit and implicit task constraints embedded in this kinesthetic demonstration can be recorded by many commercial robots such as UR robots in the joint space $\mathcal{J}$ as a time sequence of joint angles $\boldsymbol{\varphi}_{rec}$ as:

$$\boldsymbol{\varphi}_{rec} = \{\boldsymbol{\varphi}(1), \boldsymbol{\varphi}(2), \dots, \boldsymbol{\varphi}(m)\} \tag{11}$$

where each $\boldsymbol{\varphi}(i) = [\varphi_1(i), \varphi_2(i), \dots, \varphi_r(i)]^T, i = 1,2, \dots, m$ is the vector representing joint angles of the manipulator, $r$ is the degrees of freedom (DOF) of the manipulator, the order of $i$ is the time sequence of the configurations reached during the motion. Using the forward kinematics mapping $\mathcal{FK}: \mathcal{J} \to SE(3)$, the corresponding human demonstration in the task space will be obtained as $\boldsymbol{DP} = \{\boldsymbol{D}_1, \boldsymbol{D}_2, \dots, \boldsymbol{D}_m\}$, in which each $\boldsymbol{D}_i$ is a dual quaternion denoted as $\boldsymbol{D}_i = \boldsymbol{d}_{ri} + \frac{1}{2}\epsilon\boldsymbol{d}_{ti} \otimes \boldsymbol{d}_{ri}$ , $i = 1,2, \dots, m$, according to Eqn. (7). Therefore, $\boldsymbol{DP}$ represents the sequence of configurations of how a task is performed in $SE(3)$.

From the human demonstration $\boldsymbol{DP} = \{\boldsymbol{D}_1, \boldsymbol{D}_2, \dots, \boldsymbol{D}_m\}$, we compute the relative motion with respect to the final end effector configuration. Using the dual quaternion representation, the transformation $\delta_i$ between the final configuration (denoted by $\boldsymbol{D}_m$) and every other configuration $\boldsymbol{D}_i$ is:

$$\delta_i = \boldsymbol{D}_i^* \otimes \boldsymbol{D}_m \ , i = 1,2, \dots, m-1 \tag{12}$$

where $\boldsymbol{D}_i^*$ is the conjugate dual quaternion of $\boldsymbol{D}_i$ and $\otimes$ is the dual quaternion product. It is noticed that all implicit task constraints are embedded in the sequence of $\delta_i$ during the motion, which is referred to as the feature of the human demonstration. As such, the feature of the $i^{th}$ human demonstration in the task space can then be described as:

$$\boldsymbol{HD}_i = \{\delta_1, \delta_2, \dots, \delta_{m-1}\} \tag{13}$$

Therefore, the library consisting of $h$ human demonstrated features in the task space can be denoted as:

$$\boldsymbol{LB} = \{\boldsymbol{HD}_1, \boldsymbol{HD}_2, \dots, \boldsymbol{HD}_h\} \tag{14}$$

Based on the definition of the manipulation task and the library of human demonstrations, the problem studied in this paper can be described as follows: *Given a library of human demonstrations* $\boldsymbol{LB} = \{\boldsymbol{HD}_1, \boldsymbol{HD}_2, \dots, \boldsymbol{HD}_h\}$ *and a new task* $\boldsymbol{TK} = \{con_1, con_2, \dots, con_n\}$ *in the task space* $SE(3)$, *develop a method to find motion plan* $\boldsymbol{MP}$ *in the joint space* $\mathcal{J}$ *for the new task by learning from human demonstrated features in* $\boldsymbol{LB}$ *such that the explicit task space constraints specified in each* $con_i$ *in* $\boldsymbol{TK}$ *is satisfied. If no* $\boldsymbol{MP}$ *can be found, then a request for additional human demonstrations is made to satisfy all task-relevant constraints in* $\boldsymbol{TK}$.

*3.3. Build criteria for selecting appropriate human demonstration for the new task*

In a previous work (Laha, Rao, et al., 2021), we demonstrate a method of learning by demonstration, where the robot is shown the exact demonstration that needs to be learned from. In this paper, we use a more advanced and realistic scenario that the human demonstrations are abstracted by features $\boldsymbol{HD}_i$ and stored in a library $\boldsymbol{LB}$. Therefore, when performing a task specified by $\boldsymbol{TK}$, we first need to identify the appropriate one or more $\boldsymbol{HD}_i$ to be learned from.

Let $\boldsymbol{tk}_s \subseteq \boldsymbol{TK}$ be a subset or the entire new task $\boldsymbol{TK}$ as:

$$\boldsymbol{tk}_s = \{con_i, con_{i+1}, \dots, con_j\}, \quad i \geq 1, i \leq j \leq n \tag{15}$$

we need to determine a criterion that can facilitate the comparison of $\boldsymbol{tk}_s$ and a human demonstration $\boldsymbol{HD}_i$. Since $\boldsymbol{HD}_i$ represents the relative motion with respect to the final end effector configuration, in order to compare $\boldsymbol{HD}_i$ with $\boldsymbol{tk}_s$, the relative motion with respect to the final configuration in task $\boldsymbol{tk}_s$ needs to be obtained in a similar way to that of $\boldsymbol{HD}_i$. Based on Eqn. (7), the $i^{th}$ configuration of the task $\boldsymbol{tk}_s$ can be written as:

$$\boldsymbol{D}_i^{tk_s} = \boldsymbol{d}_{ri}^{tk_s} + \frac{1}{2}\epsilon \boldsymbol{d}_{ti}^{tk_s} \otimes \boldsymbol{d}_{ri}^{tk_s}, i = 1,2, \dots, n \tag{16}$$

where $\boldsymbol{d}_{ti}^{tk_s} = (0, \boldsymbol{P}_i)$, $\boldsymbol{P}_i$ is the position of the $i^{th}$ critical configuration, $\boldsymbol{d}_{ri}^{tk_s}$ is a unit quaternion that can be expressed as:

$$\boldsymbol{d}_{ri}^{tk_s} = \begin{matrix} \cos\left(\frac{\theta_{i3}}{2}\right)\cos\left(\frac{\theta_{i2}}{2}\right)\cos\left(\frac{\theta_{i1}}{2}\right) + \sin\left(\frac{\theta_{i3}}{2}\right)\sin\left(\frac{\theta_{i2}}{2}\right)\sin\left(\frac{\theta_{i1}}{2}\right) + \\ \left(\sin\left(\frac{\theta_{i3}}{2}\right)\cos\left(\frac{\theta_{i2}}{2}\right)\cos\left(\frac{\theta_{i1}}{2}\right) - \cos\left(\frac{\theta_{i3}}{2}\right)\sin\left(\frac{\theta_{i2}}{2}\right)\sin\left(\frac{\theta_{i1}}{2}\right)\right)\hat{\boldsymbol{\imath}} + \\ \left(\cos\left(\frac{\theta_{i3}}{2}\right)\sin\left(\frac{\theta_{i2}}{2}\right)\cos\left(\frac{\theta_{i1}}{2}\right) + \sin\left(\frac{\theta_{i3}}{2}\right)\cos\left(\frac{\theta_{i2}}{2}\right)\sin\left(\frac{\theta_{i1}}{2}\right)\right)\hat{\boldsymbol{\jmath}} + \\ \left(\cos\left(\frac{\theta_{i3}}{2}\right)\cos\left(\frac{\theta_{i2}}{2}\right)\sin\left(\frac{\theta_{i1}}{2}\right) - \sin\left(\frac{\theta_{i3}}{2}\right)\sin\left(\frac{\theta_{i2}}{2}\right)\cos\left(\frac{\theta_{i1}}{2}\right)\right)\hat{\boldsymbol{k}} \end{matrix} \tag{17}$$

where $\boldsymbol{\theta}_i = (\theta_{i1}, \theta_{i2}, \theta_{i3})$ is the orientation of the $i^{th}$ critical configuration of the end-effector. Therefore, the transformation, $\delta_i^{tk_s}$, between the last critical configuration $con_n$ and any other critical configuration $con_i$ is defined as:

$$\delta_i^{tk_s} = \boldsymbol{D}_i^{tk_s^*} \otimes \boldsymbol{D}_n^{tk_s}, i = 1, \dots, n-1 \tag{18}$$

As such, the feature of the task $\boldsymbol{tk}_s$ can be represented as $\boldsymbol{tk}_s^{\delta} = \{\delta_1^{tk_s}, \delta_2^{tk_s}, \dots, \delta_{n-1}^{tk_s}\}$.

Since both $\delta_i$ and $\delta_j^{tk_s}$ are represented in dual quaternions as $\delta_i = \boldsymbol{d}_r^{\delta_i} + \frac{1}{2}\epsilon \boldsymbol{d}_t^{\delta_i} \otimes \boldsymbol{d}_r^{\delta_i}$ and $\delta_j^{tk_s} = \boldsymbol{d}_r^{\delta_j^{tk_s}} + \frac{1}{2}\epsilon \boldsymbol{d}_t^{\delta_j^{tk_s}} \otimes \boldsymbol{d}_r^{\delta_j^{tk_s}}$ based on Eqn. (7), the similarities between them are evaluated by the closeness/difference with respect to both rotation and translation. The closeness of their rotation can be evaluated using Euclidean distance as:

$$\alpha\left(\delta_i, \delta_j^{tk_s}\right) = \min\left\{\left\|\boldsymbol{d}_r^{\delta_i} - \boldsymbol{d}_r^{\delta_j^{tk_s}}\right\|, \left\|\boldsymbol{d}_r^{\delta_i} + \boldsymbol{d}_r^{\delta_j^{tk_s}}\right\|\right\} \tag{19}$$

The difference between the translation direction of the $\delta_i$ and $\delta_j^{tk_s}$ is evaluated as the dot product of the normalized translation vector as:

$$\beta\left(\delta_i, \delta_j^{tk_s}\right) = \frac{\boldsymbol{d}_t^{\delta_i}}{\left|\boldsymbol{d}_t^{\delta_i}\right|} \cdot \frac{\boldsymbol{d}_t^{\delta_j^{tk_s}}}{\left|\boldsymbol{d}_t^{\delta_j^{tk_s}}\right|} \tag{20}$$

Given tolerances $\Delta_\alpha$ and $\Delta_\beta$, if $\alpha\left(\delta_i, \delta_j^{tk_s}\right) \leq \Delta_\alpha$ and $\beta\left(\delta_i, \delta_j^{tk_s}\right) \geq \Delta_\beta$, then these two transformations are referred to as being "semantically similar".

**Remark 2**. According to a previous work (Laha, Rao, et al., 2021), if two transformations are sufficiently close within a certain tolerance (the practical value is about 10 degrees according to (Laha, Rao, et al., 2021)) in the task space, the corresponding solutions in the joint space using inverse kinematics can be uniquely determined by a small increment in joint angles. Since the human demonstration guarantees the feasible solution in the joint space without violating any joint limits, if the new task is "semantically similar" to a certain human demonstration based on Eqn. (19) and Eqn. (20), by learning from that human demonstration, the solution in the joint space can be ensured. Therefore, a criterion of semantic similarity between a task and a demonstration can be defined.

**Definition 1.** Let $p$ be the number of critical configurations in a task feature $\boldsymbol{tk}_s^{\delta}$, the task feature $\boldsymbol{tk}_s^{\delta}$ and a human demonstrated feature $\boldsymbol{HD}_k$ are semantically similar, denoted as $\boldsymbol{tk}_s^{\delta} \propto \boldsymbol{HD}_k$, if the same number of $p$ configurations on $\boldsymbol{HD}_k$ can be allocated such that the following criterion is satisfied:

$$\forall \delta_j^{tk_s} \in \boldsymbol{tk}_s^{\delta}, j = 1, \dots, p, \exists \delta_l \in \boldsymbol{HD}_k \;\; \alpha\left(\delta_l, \delta_j^{tk_s}\right) \leq \Delta_\alpha \text{ and } \beta\left(\delta_l, \delta_j^{tk_s}\right) \geq \Delta_\beta \tag{21}$$

where $l = i, \dots, m$ and $m$ is the last configuration in $\boldsymbol{HD}_k$.

For example, for the same task $\boldsymbol{TK}$ shown in Fig. 3, four critical configurations can be specified in Table 2 and corresponding features can be derived in Table 3 using Eqn. (18). Suppose there exists a library $\boldsymbol{LB}$ that includes features $\boldsymbol{HD}_1$ and $\boldsymbol{HD}_2$ summarized in Table 4, we want to compare these features with the feature of the task $\boldsymbol{TK}$. From Table 3 and Table 4, we find that, for each $\delta_i^{\boldsymbol{TK}}, i = 1,2$, we can find a $\delta_i$ in $\boldsymbol{HD_1}$, such that $\alpha\left(\delta_i, \delta_i^{\boldsymbol{TK}}\right) = 0$ and $\beta\left(\delta_i, \delta_i^{\boldsymbol{TK}}\right) \geq 0$. For $\delta_3^{\boldsymbol{TK}}$, we find $\delta_1$ in $\boldsymbol{HD}_2$, such that $\alpha(\delta_1, \delta_3^{\boldsymbol{TK}}) = 0$ and $\beta(\delta_1, \delta_3^{\boldsymbol{TK}}) \geq 0$. Given $\Delta_\alpha = 0.5$ and $\Delta_\beta = 0$, based on Definition 1, the feature of the segment between $con_1$ and $con_3$ for task $\boldsymbol{TK}$ is semantically similar to that of $\boldsymbol{HD}_1$ and the feature of the segment between $con_3$ and $con_4$ for task $\boldsymbol{TK}$ is semantically similar to that of $\boldsymbol{HD}_2$.

**Table 2**. Critical Configurations of $\boldsymbol{TK}$

| | $con_1$ | $con_2$ | $con_3$ | $con_4$ |
|---|---|---|---|---|
| $\boldsymbol{P}$ | (-0.2,0,0.6) | (-0.3,0,0.6) | (-0.4,0,0.6) | (-0.5,0,0.6) |
| $\boldsymbol{\theta}$ | (0,0,0) | (0, 0,0) | (0,0,0) | $(0,0,-\pi/2)$ |

**Table 3**. Features of $\boldsymbol{TK}$

| | $\delta_1^{\boldsymbol{TK}}$ | $\delta_2^{\boldsymbol{TK}}$ | $\delta_3^{\boldsymbol{TK}}$ |
|---|---|---|---|
| $\boldsymbol{d}_t^{\delta^{TK}}$ | (0,1,0,0) | (0,1,0,0) | (0,1,0,0) |
| $\boldsymbol{d}_r^{\delta^{TK}}$ | (1,0,0,0) | (1,0,0,0) | $(0.7,0,0,-0.7)$ |

**Table 4**. Features of Human Demonstrations

| | | $\boldsymbol{HD}_1$ | $\boldsymbol{HD}_2$ |
|---|---|---|---|
| $\boldsymbol{\delta}_1$ | $\boldsymbol{d}_t^{\delta_1}$ | (0,1,0,0) | (0,0.7,0,-0.7) |
| | $\boldsymbol{d}_r^{\delta_1}$ | (0.7,0,0,-0.7) | (0.7,0,0.7,0) |
| $\boldsymbol{\delta}_2$ | $\boldsymbol{d}_t^{\delta_2}$ | (0,1,0,0) | |
| | $\boldsymbol{d}_r^{\delta_2}$ | (0.9,0,0,-0.4) | |

*3.4. Map the feature of the selected human demonstration to the new task*

In a previous work (Laha, Rao, et al., 2021), to generate the point-to-point motion plan for similar task instances, an imitated path is built by transmitting the human demonstration in the task space to the new goal position and the ScLERP (Kavan et al., 2006) is used to blend any current configuration of the end-effector into the imitated path. However, this method only considers the goal position of the new task instance as the explicit new task constraints and the implicit task constraints may not be satisfied during the blend in motion.

Therefore, the method has the limit in performing similar tasks in close adjacent areas of human demonstrations. In this paper, we adopt a different approach by using the selected feature $\boldsymbol{HD}_k$ based on the criterion in Eqn. (21), and mapping the feature to a new task or a segment of the new task.

Suppose $\boldsymbol{HD}_k$ and a new task $\boldsymbol{tk}_s$ are semantically similar based on Definition 1, i.e., $\boldsymbol{tk}_s \propto \boldsymbol{HD}_k$, we will use mapping $\boldsymbol{mp}_k^s\colon \boldsymbol{HD}_k \to \boldsymbol{tk}_s$ , such that to finish the new task $\boldsymbol{tk}_s$, the robot can learn from $\boldsymbol{HD}_k$. Let $\boldsymbol{HD}_k = \{\delta_1, \delta_2, \dots, \delta_{m-1}\}$, $\boldsymbol{tk}_s^{\delta} = \{\delta_1^{\boldsymbol{tk}_s}, \delta_2^{\boldsymbol{tk}_s}, \dots, \delta_{n-1}^{\boldsymbol{tk}_s}\}$, the first step is to align the translation vector $\boldsymbol{d}_t^{\delta_1}$ in $\boldsymbol{HD}_k$ to the translation vector $\boldsymbol{d}_t^{\delta_1^{\boldsymbol{tk}_s}}$ in $\boldsymbol{tk}_s^{\delta}$. Based on Eqn. (9), a unit quaternion $\boldsymbol{d}_{\boldsymbol{HD}_k}^{\boldsymbol{tk}_s}$ representing rotating the vector $\boldsymbol{d}_t^{\delta_1}$ to the vector $\boldsymbol{d}_t^{\delta_1^{\boldsymbol{tk}_s}}$ is defined as:

$$\boldsymbol{d}_{\boldsymbol{HD}_k}^{\boldsymbol{tk}_s} = \cos\left(\frac{\phi_{\boldsymbol{HD}_k}^{\boldsymbol{tk}_s}}{2}\right) + \widehat{\boldsymbol{n}}_{\boldsymbol{HD}_k}^{\boldsymbol{tk}_s} \sin\left(\frac{\phi_{\boldsymbol{HD}_k}^{\boldsymbol{tk}_s}}{2}\right) \tag{22}$$

where $\widehat{\boldsymbol{n}}_{\boldsymbol{HD}_k}^{\boldsymbol{tk}_s} = \frac{\boldsymbol{d}_t^{\delta_1} \times \boldsymbol{d}_t^{\delta_1^{\boldsymbol{tk}_s}}}{\left|\boldsymbol{d}_t^{\delta_1} \times \boldsymbol{d}_t^{\delta_1^{\boldsymbol{tk}_s}}\right|}$ and $\phi_{\boldsymbol{HD}_k}^{\boldsymbol{tk}_s} = \cos^{-1}\frac{\boldsymbol{d}_t^{\delta_1} \cdot \boldsymbol{d}_t^{\delta_1^{\boldsymbol{tk}_s}}}{\left|\boldsymbol{d}_t^{\delta_1}\right|\left|\boldsymbol{d}_t^{\delta_1^{\boldsymbol{tk}_s}}\right|}$. Since $\boldsymbol{HD}_k$ and $\boldsymbol{tk}_s$ are semantically similar, each $\boldsymbol{d}_t^{\delta_1}$ of $\delta_i$ in $\boldsymbol{HD}_k, i = 1,2,\dots,m-1$ can be rotated using the quaternion sandwich $\boldsymbol{d}_{\boldsymbol{HD}_k}^{\boldsymbol{tk}_s} \boldsymbol{d}_t^{\delta_i} \boldsymbol{d}_{\boldsymbol{HD}_k}^{\boldsymbol{tk}_s *}$. Then, we can scale the vector $\boldsymbol{d}_{\boldsymbol{HD}_k}^{\boldsymbol{tk}_s} \boldsymbol{d}_t^{\delta_i} \boldsymbol{d}_{\boldsymbol{HD}_k}^{\boldsymbol{tk}_s *}$ with $\frac{\left|\boldsymbol{d}_t^{\delta_1}\right|}{\left|\boldsymbol{d}_t^{\delta_1^{\boldsymbol{tk}_s}}\right|}$. The final translation quaternion $\boldsymbol{d}_{ti}^{\boldsymbol{mp}_k^s}$ that maps the relative translation $\boldsymbol{d}_t^{\delta_i}$ in $\boldsymbol{HD}_k$ to $\boldsymbol{tk}_s$ can be obtained as:

$$\boldsymbol{d}_t^{\boldsymbol{mp}_k^s} = \left(0, \frac{\left|\boldsymbol{d}_t^{\delta_1}\right|}{\left|\boldsymbol{d}_t^{\delta_1^{\boldsymbol{tk}_s}}\right|} \boldsymbol{d}_{\boldsymbol{HD}_k}^{\boldsymbol{tk}_s} \boldsymbol{d}_t^{\delta_i} \boldsymbol{d}_{\boldsymbol{HD}_k}^{\boldsymbol{tk}_s *}\right), i = 1,2,\dots.,m-1 \tag{23}$$

Since the detailed transformation between $con_j$ and $con_{j+1}$ are unknown and unspecified, and more importantly, due to the rational in Remark 2, we can just use all the intermediate transformation in $\delta_i$ for the transformation between $con_j$ and $con_{j+1}$. Let $\boldsymbol{d}_{ri}^{\boldsymbol{mp}_k^s} = \boldsymbol{d}_r^{\delta_i}, i = 1,2,\dots.,m-1$, the mapping $\boldsymbol{mp}_k^s$ can finally be derived as:

$$\boldsymbol{mp}_k^s = \boldsymbol{d}_r^{\boldsymbol{mp}_k^s} + \frac{1}{2} \boldsymbol{d}_t^{\boldsymbol{mp}_k^s} \otimes \boldsymbol{d}_r^{\boldsymbol{mp}_k^s} \tag{24}$$

Where $\boldsymbol{d}_r^{\boldsymbol{mp}_k^s} = \{\boldsymbol{d}_{r1}^{\boldsymbol{mp}_k^s}, \boldsymbol{d}_{r2}^{\boldsymbol{mp}_k^s}, \dots, \boldsymbol{d}_{rm-1}^{\boldsymbol{mp}_k^s}\}$, $\boldsymbol{d}_t^{\boldsymbol{mp}_k^s} = \{\boldsymbol{d}_{t1}^{\boldsymbol{mp}_k^s}, \boldsymbol{d}_{t2}^{\boldsymbol{mp}_k^s}, \dots, \boldsymbol{d}_{tm-1}^{\boldsymbol{mp}_k^s}\}$. Let $\boldsymbol{TK} = \{tk_1, tk_2, \dots, tk_p\}$, the final motion plan $\boldsymbol{MP}$ in task space for $\boldsymbol{TK}$ can be determined by

$$\boldsymbol{MP} = \boldsymbol{D}_n^{tk_s} \otimes \boldsymbol{mp}_k^{s*}, s = 1,2, \dots, p \quad (25)$$

If the requirements of all $\boldsymbol{tk}_s$ of a new task $\boldsymbol{TK}$ can be covered by features of human demonstrations in $\boldsymbol{LB}$, then it is theoretically possible that there exists a set $\{\boldsymbol{HD_i}, \dots, \boldsymbol{HD_l}\} \subseteq \boldsymbol{LB}$ **,** from which the robot can learn to finish $\boldsymbol{TK}$ by using mapping. This is made more rigorous in Theorem 1.

**Theorem 1.** *Given a library of human demonstrations* $\boldsymbol{LB} = \{\boldsymbol{HD}_1, \boldsymbol{HD}_2, \dots, \boldsymbol{HD}_h\}$ *and a new task* $\boldsymbol{TK} = \{\boldsymbol{tk}_1, \boldsymbol{tk}_2, , \dots, \boldsymbol{tk}_p\}$ *in SE(3), it is always possible to find a set* $\boldsymbol{HD} = \{\boldsymbol{HD}_i, \dots, \boldsymbol{HD}_l\} \subseteq \boldsymbol{LB}$ *such that* $\forall\, \boldsymbol{tk}_s \subseteq \boldsymbol{TK}, s = 1,2, \dots, p$, $\exists \boldsymbol{HD}_k \in \boldsymbol{HD}$ *such that* $\boldsymbol{tk_s} \propto \boldsymbol{HD_k}$, *if and only if* $\forall\, \boldsymbol{tk}_s \subseteq \boldsymbol{TK}, s = 1,2, \dots, p$, $\exists \boldsymbol{HD}_j \in \boldsymbol{LB}$ *such that* $\boldsymbol{tk_s} \propto \boldsymbol{HD}_j$

*Proof*

If $\forall\, \boldsymbol{tk}_s \subseteq \boldsymbol{TK}, s = 1,2, \dots, p$, $\exists \boldsymbol{HD}_j \in \boldsymbol{LB}$ *such that* $\boldsymbol{tk}_s \propto \boldsymbol{HD}_j$, then we can always find a set that includes $\boldsymbol{HD}_j$, i.e., $\boldsymbol{HD} = \{\boldsymbol{HD}_i, \boldsymbol{HD}_j \dots, \boldsymbol{HD_l}\} \subseteq \boldsymbol{LB}$ that satisfy the above condition. To prove the sufficiency, suppose we can find a a set $\boldsymbol{HD} = \{\boldsymbol{HD}_i \dots, \boldsymbol{HD_l}\} \subseteq \boldsymbol{LB}$**,** that $\forall\, \boldsymbol{tk}_s \subseteq \boldsymbol{TK}, s = 1,2, \dots, p$, $\exists \boldsymbol{HD}_k \in \boldsymbol{HD}$ such that $\boldsymbol{tk_s} \propto \boldsymbol{HD}_k$. Since $\boldsymbol{HD}_k \in \boldsymbol{HD} = \{\boldsymbol{HD}_i \dots, \boldsymbol{HD}_l\} \subseteq \boldsymbol{LB}$, then sufficient condition must be true. ■

In the next section, we will discuss how to find a set $\boldsymbol{HD} = \{\boldsymbol{HD}_i, \boldsymbol{HD}_k, \dots, \boldsymbol{HD}_l\}$, meaning find the motion planning by learning from $\boldsymbol{HD}$ in the task space. This problem is formulated into a Markov Decision Process (MDP) problem and solved with Q-learning.

## 4. Obtaining the Optimal Motion Plan Trough Reinforcement Learning

To find a set of appropriate features in $\boldsymbol{LB}$ that can satisfy all task-relevant constraints in $\boldsymbol{TK}$, one has to go through all subsets $\boldsymbol{tk}_s$ of $\boldsymbol{TK}$ and evaluate each $\boldsymbol{HD}_k$ in the library $\boldsymbol{LB}$, which is an NP-hard problem (Bovet et al., 1994). Assume that there are $x$ features stored in the library $\boldsymbol{LB}$ and $y$ subsets of the new task $TK$, the computational complexity for searching exhaustively is $O(x^y)$, which would be huge if the number of human demonstrations and the constraints of the new task is large. For example, the library of "Assemble a height-adjustable desk" have 10 features including flipping, twisting, passing, etc. A new task "Place the screw into the assembly hole and fasten the screw" can have 20 subsets (such as passing horizontally to certain position, then change direction to another position, etc.). In this case, the computational complexity for exhaustive search is $O(10^{20})$. Therefore, we formulate the problem into an MDP and solve the problem using a model-free reinforcement learning (RL) algorithm.

*4.1. MDP Formulation of the Problem*

The most common framework for RL is MDP, which is a stochastic process that models the sequential decision making in uncertain environments. There are three components in an MDP, including state $s$, action $a$ and reward function $r$. In a RL framework, an agent's objective is to find a policy $\pi$ so as to maximize the sum of discounted expected rewards

$$v(s,\pi) = \sum_{t=0}^{T} \gamma^t E(r_t|\pi, s_0 = s) \tag{26}$$

where $v(s,\pi)$ is the value for state $s$ under the policy $\pi$. Here $\pi = (\pi_0, \dots, \pi_t, \dots)$ is defined over the entire process. The standard solution to the problem above is through the $Bellman$ equation:

$$v(s,\pi^*) = \max_a [r(s,a) + \gamma \sum_{s'} p(s'|s,a) v(s',\pi^*)] \tag{27}$$

where $r(s,a)$ is the reward for taking action $a$ at state $s$, $s'$ is the next state, and $p(s'|s,a)$ is the probability of transiting to state $s'$ after taking action $a$ in state $s$. A solution $\pi^*$ that satisfies the above equation is guaranteed to be an optimal policy. Before we can apply RL algorithms to obtaining the ultimate motion planning policy $\pi^*$, we need to first properly define the three key components $s_t$, $a_t$ and $r_t$ at time step $t$.

The state $s_t$ is defined as:

$$s_t = [\boldsymbol{CF}_t, \boldsymbol{tk}_t] \tag{28}$$

where $\boldsymbol{CF}_t$ is the current configuration of the end-effector at $t$, $\boldsymbol{tk}_t = \{con_j, con_{j+1}, \dots, con_n\}$ is the subset of $\boldsymbol{TK}$ containing task constraints that the robot is going to satisfy.

The action $a_t$ can be defined as:

$$a_t = [a_1(t), a_2(t), \dots, a_h(t)] \tag{29}$$

where each $a_i(t)$ is to identify a subsequent critical point $con_k$ that $\boldsymbol{tk}'_t = \{con_j, \dots, con_k\} \subseteq \{con_j, con_{j+1}, \dots, con_n\}$, such that $\boldsymbol{tk}'_t \propto \boldsymbol{HD}_i$. Then $a_i(t)$ will take the index value of $k$. Therefore, $a_i(t)$ is defined as:

$$a_i(t) = \begin{cases} k, & \text{if } \boldsymbol{tk}'_t \propto \boldsymbol{HD}_i \\ 0, & \text{otherwise} \end{cases} \tag{30}$$

The feature of $\boldsymbol{tk}'_t$ can be written as a set of $\delta_i^{\boldsymbol{tk}'_t}, i = j, \dots, k-1$, which is obtained through Eqn. (18).

To evaluate the action at $t$, the reward function $r_t$ is defined as:

$$r_t = \begin{cases} -\sum_{l=j}^{k} \alpha\left(\delta_i, \delta_l^{\boldsymbol{tk}_t'}\right), & \text{if } \boldsymbol{tk}_t' \propto \boldsymbol{HD_i} \\ -\infty, & \text{otherwise} \end{cases} \tag{31}$$

*4.2. Applying Q-learning to Obtain the Optimal Motion Planning Policy*

In order to obtain the optimal policy $\pi^*$, various algorithms have been proposed in the past, among which Q-learning is one of the most widely used algorithms (Sutton & Barto, 2018). The basic idea of Q-learning is that we can define a function $Q$:

$$Q(s,a) = r(s,a) + \gamma \sum_{s' \in S} p(s'|s,a) v(s',\pi) \tag{32}$$

such that $v(s,\pi^*) = \max_a Q^*(s,a)$. If we know $Q^*(s,a)$, then the optimal policy $\pi^*$ can be found by simply identifying the action that maximizes $Q^*(s,a)$ under the state $s$. Starting with arbitrary initial values of $Q(s,a)$, the updating procedure of Q-learning is:

$$Q_{t+1}(s_t,a_t) = (1-\alpha_t)Q_t(s_t,a_t) + \alpha_t[r_t + \gamma \max_a Q_t(s_{t+1},a_t)] \tag{33}$$

where $\alpha_t \in [0,1)$ is the learning rate and $\gamma \in (0,1)$ is the discount factor. The training process is shown in **Algorithm 1**. After the training, the ultimate policy $\pi^*$ is determined as:

$$\pi^*(a|s) = \begin{cases} 1, & \text{if } a = \arg\max_{a' \in A(s)} \{Q(s,a')\} \\ 0, & \text{otherwise} \end{cases} \tag{34}$$

where $A(s)$ is the set of all legal actions at state $s$ in the Q-table $Q(s,a)$. The final motion plan $\boldsymbol{MP}$ is generated following the **Algorithm 2.** The overall framework of robot motion planning based on Q-learning by learning from human demonstrations is illustrated in Fig. 4.

**Algorithm 1 Training of the RL-based Motion Planner in $SE(3)$**

---

**Procedure 1** Mapping Features of the Human Demonstration to the New Task

**Input:** $\boldsymbol{tk}_s$, $\boldsymbol{HD}_k$

**Output:** $\boldsymbol{mp}_k^s$

Initialize $\boldsymbol{mp}_k^s$ with zeros

Compute each $\delta_i^{tk_s}$ using Eqn. (18)

Compute the rotation quaternion $\boldsymbol{d}_{\boldsymbol{HD}_k}^{\boldsymbol{tk}_s}$ using Eqn. (22)

**For** $k = 1,2,\dots,m-1$ **do**

  Update each $\boldsymbol{d}_{ti}^{\boldsymbol{mp}_k^s}$ using Eqn. (23)

  $\boldsymbol{d}_{ri}^{\boldsymbol{mp}_k^s} \leftarrow \boldsymbol{d}_r^{\delta_i}$

**End For**

Compute $\boldsymbol{mp}_k^s$ using Eqn. (24)

Output $\boldsymbol{mp}_k^s$

**End Procedure**

**Procedure 2** Training Process of Q-learning

**Input**: $\boldsymbol{TK}$, $\boldsymbol{LB}$, $\epsilon, \gamma$

Output: $Q(s, a)$

Initialize $Q(s, a)$ randomly

Initialize $t = 0$

Initialize $s_0$ with $\boldsymbol{CF}_0 = \boldsymbol{D}_1^{tk_1}$ according to Eqn. (16) and $\boldsymbol{tk}_0 = \boldsymbol{TK}$

**For** $episode = 0,1, \dots ,100$ **do**

  **While** the last $con_n$ of $\boldsymbol{TK}$ is not reached **do**

    Choose $a_t$ using policy derived from $Q(s, a)$ (e.g., $\epsilon$-greedy)

    Take action $a_t$, observe $r_t$

    **If** $\boldsymbol{tk}_t' \propto \boldsymbol{HD_i}$ based on Definition 1

      $Q(s_t, a_t) \leftarrow Q(s_t, a_t) + \alpha \left[r_t + \gamma \max_{a'} Q(s_{t+1}, a') - Q(s_t, a_t)\right]$

      $i \leftarrow$ index of the non-zero term of $a_t$

      Invoke Procedure 1 with $\boldsymbol{HD}_i$ and $\boldsymbol{tk}_t'$ as inputs

      $\boldsymbol{CF}_{t+1} \leftarrow \boldsymbol{D}_n^{tk_t'}$

      $\boldsymbol{tk}_{t+1} \leftarrow \{con_{a_i(t)}, \dots , con_n\}$

      $s_t \leftarrow s_{t+1}$

      $t \leftarrow t + 1$

    **Else**

      $Q(s_t, a_t) \leftarrow Q(s_t, a_t) + \alpha \left[r_t + \gamma \max_{a'} Q(s_{t+1}, a') - Q(s_t, a_t)\right]$

      **Break**

    **End if**

  **End While**

**End For**

Output $Q(s, a)$

**End Procedure**

**Algorithm 2 Generating Motion Plan in $\boldsymbol{SE}(\boldsymbol{3})$**

**Input:** $\boldsymbol{TK}$, $\boldsymbol{LB}$, $\epsilon, \gamma, Q(s, a)$

**Output:** $\boldsymbol{MP}$

Initialize $s_0$ with $\boldsymbol{CF}_0 = \boldsymbol{D}_1^{tk_1}$ according to Eqn. (16) and $\boldsymbol{tk}_0 = \boldsymbol{TK}$

**For** $t = 0,1, \dots , T$ **do**

  Find legal action list $A(s_t)$ from $Q(s_t, a) \rightarrow a \in A(s_t)$

  Find the optimal action as $a_t = \arg \max_{a \in A(s_t)} Q(s_t, a)$

Map $\boldsymbol{HD}_i$ to $\boldsymbol{tk}_t'$ according to $a_t$

$MP_t \leftarrow \boldsymbol{D}_n^{tk_t'} \otimes \boldsymbol{mp}_i^{t*}$ according to Eqn. (24)

**End For**

$\boldsymbol{MP} \leftarrow \{MP_0, MP_1, \dots, MP_T\}$

Output $\boldsymbol{MP}$

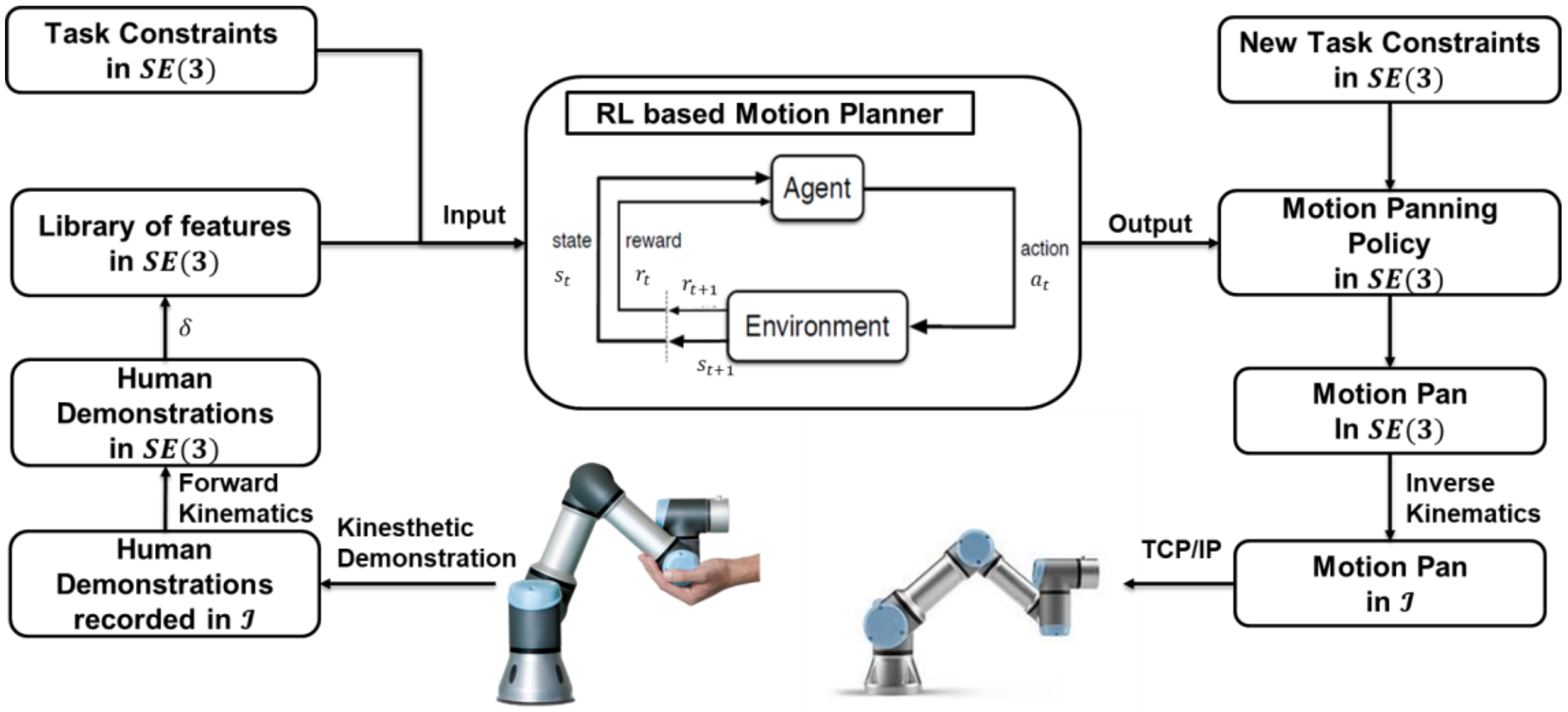

**Figure. 4** Framework of the RL based user-guided motion planning.

## 5. Computing the Motion Plan in the Joint Space

After the motion plan $\boldsymbol{MP}$ in the task space is generated, the corresponding motion plan in the joint space needs to be calculated through inverse kinematics. Based on a previous work (Laha, Rao, et al., 2021), to calculate the sequence of joint angles in the joint space is to solve the equation

$$\dot{\boldsymbol{q}} = (\mathbf{J}^s)^T(\mathbf{J}^s(\mathbf{J}^s)^T)^{-1}\mathbf{J}_2\begin{bmatrix}\dot{\boldsymbol{p}}\\ \dot{\boldsymbol{r}}\end{bmatrix} \tag{35}$$

where $\dot{\boldsymbol{q}}$ is the joint rate vector, $\boldsymbol{p}$ is the position of the end-effector, $\boldsymbol{r}$ is the quaternion representing the orientation of the end-effector, $\mathbf{J}^s$ is the spatial Jacobian and $\mathbf{J}_2 = \begin{bmatrix}\mathbf{I}_{3\times3} & 2\boldsymbol{p}\mathbf{J}_1\\ \mathbf{0}_{3\times3} & 2\mathbf{J}_1\end{bmatrix}$ where $\mathbf{J}_1$ is the matrix transformation of the spatial angular velocity of the rigid body. Let $\dot{\boldsymbol{\gamma}} = [\dot{\boldsymbol{p}}\ \dot{\boldsymbol{r}}]^T$ and $\boldsymbol{B} = (\mathbf{J}^s)^T(\mathbf{J}^s(\mathbf{J}^s)^T)^{-1}\mathbf{J}_2$, Eqn. (35) can also be written as:

$$\dot{\boldsymbol{q}} = \boldsymbol{B}\dot{\boldsymbol{\gamma}} \tag{36}$$

Using Euler time-step to discretize this equation where $h$ is a small time step,

$$\frac{\boldsymbol{q}(t+h) - \boldsymbol{q}(t)}{h} = \boldsymbol{B}\frac{\boldsymbol{\gamma}(t+h) - \boldsymbol{\gamma}(t)}{h} \tag{37}$$

or

$$\delta \boldsymbol{q} = \boldsymbol{B}\big(\boldsymbol{\gamma}(t+h) - \boldsymbol{\gamma}(t)\big) \tag{38}$$

In such way, for two consecutive configurations of the end-effector $\boldsymbol{\gamma}(t)$ and $\boldsymbol{\gamma}(t+h)$**,** the value of $\delta\boldsymbol{q}$ is determined. Therefore, starting from $\boldsymbol{q}(0)$**,** the joint angles $\boldsymbol{q}(t)$ at any time step $t$ can be obtained.

In practice, an unavoidable problem when executing motion plans in the joint space is the joint limits. In previous studies, researchers usually handle joint limits with optimization or Nullspace of the Jacobian (Flacco et al., 2015). However, in this paper, we still have limitations in handling joint limits because task constraints in $SE(3)$ always have the higher priority than joint limit constraints in $\mathcal{J}$. Since we map the feature of the normalized human demonstration to the new task in the task space, the motion plan obtained by Q-learning in $SE(3)$ cannot guarantee feasible solutions in $\mathcal{J}$ that do not violate joint limits. There're also some other methods in handling joint limits (Moe et al., 2016), we will explore these options in our future work.

**6. Case Study**

In order to validate effectiveness of the proposed method in generating motion plans for new tasks, multiple experiments are conducted on the UR5e platform. Our experiments consist of four steps: (1) Build an illustrative library of features of human demonstrations; (2) Specify new tasks in $SE(3)$; (3) Offline train the RL-based Motion Planner in $SE(3)$ and obtain motion plans in $\mathcal{J}$; (4) Execute motion plans in $\mathcal{J}$. In this case study, two performance metrices are considered: (1) The accumulated reward of the motion plan in $SE(3)$; (2) The successful rate of applying the motion plan in $\mathcal{J}$ for new tasks. From the case study, two significant results can be concluded: (1) The proposed RL-based user-guided motion planning method can benefit from sufficient knowledge in the proposed task specification scheme; (2) The proposed method is effective in combining different features of human demonstrations to generate motion plans for the new task; (3) The proposed method is effective in requesting additional human demonstrations if no features in the library are semantically similar to the new task.

### *6.1. Setting up a Library of Features of User Demonstrations*

To build a library including some common features in certain assembly and loading/unloading environment, we start with five illustrative tasks as shown in Fig. 5. The tasks are recorded in joint space $\mathcal{J}$ through kinesthetic demonstrations, including:

1) Screwing Task 1: Twist a screw driver 90 degrees clockwise;
2) Screwing Task 2: Twist a screw driver 90 degrees anti-clockwise;
3) Filling Task: Hold a cup horizontally and then turn down 90 degrees (representing certain orientation constraints);
4) Pouring Task: Hold a cup vertically and then turn up 90 degrees (representing certain orientation constraints);
5) Stacking Task: Stack one block from an initial location to a goal location;

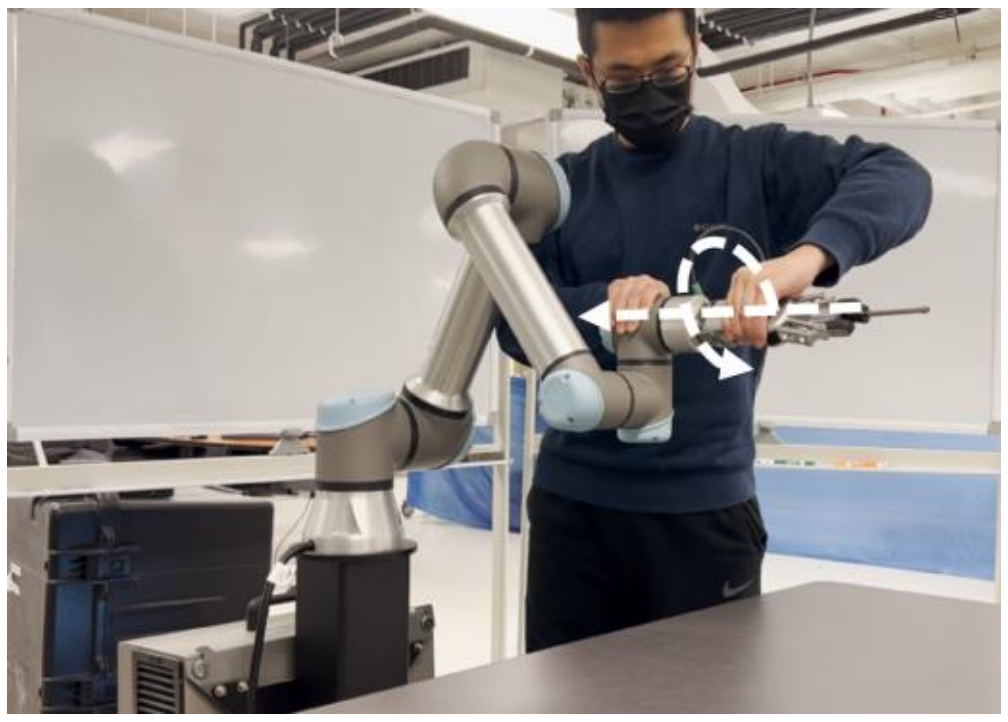

(a)

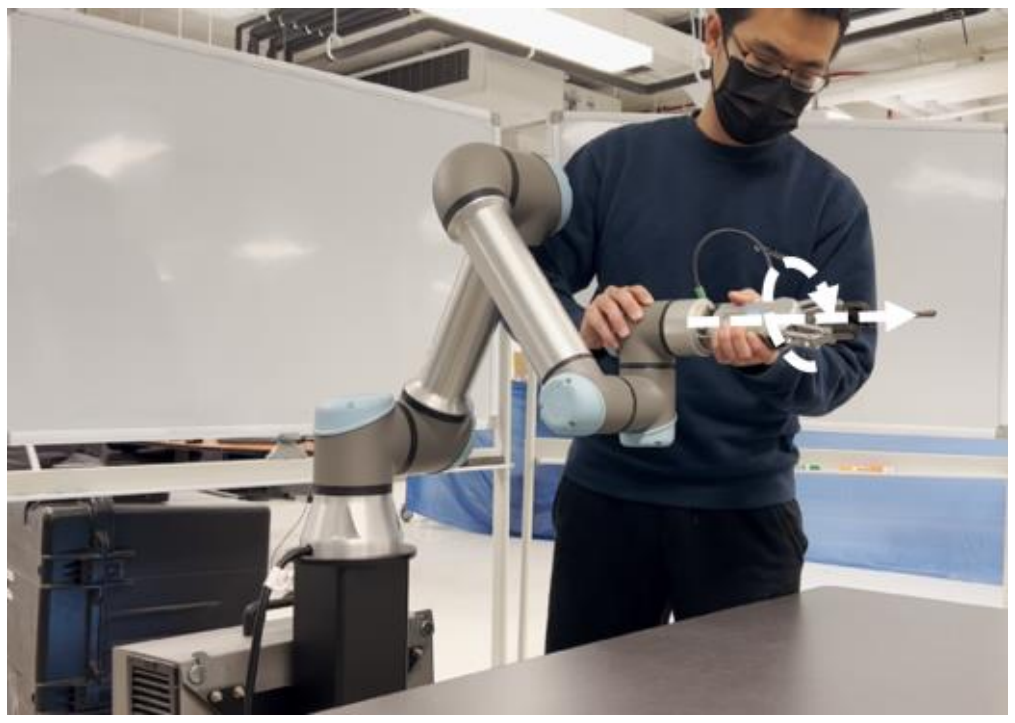

(b)

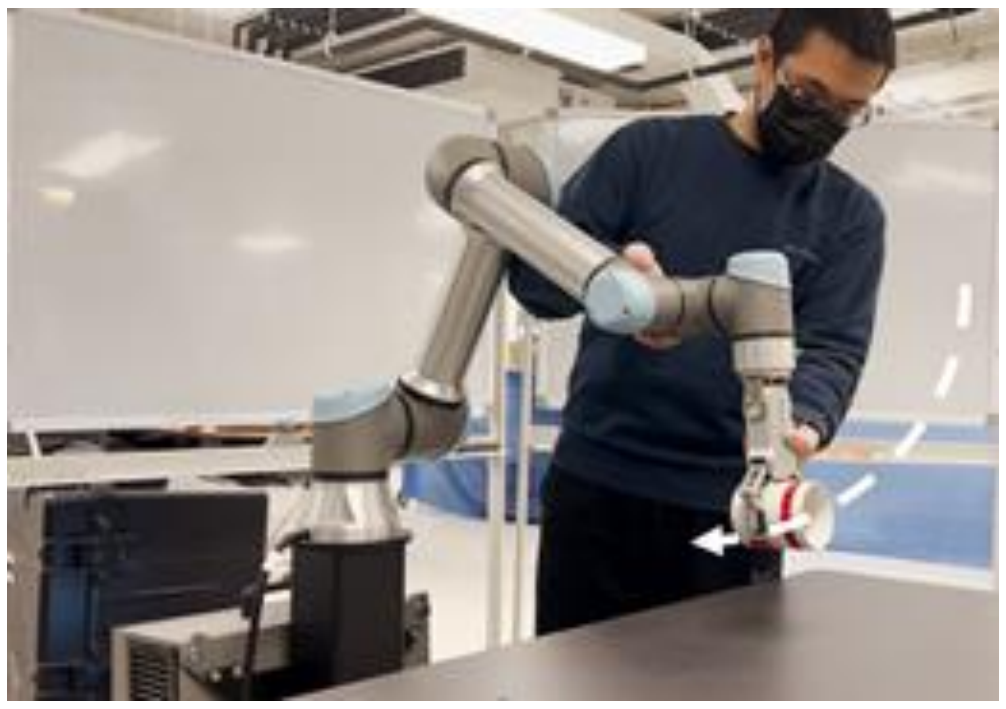

(c)

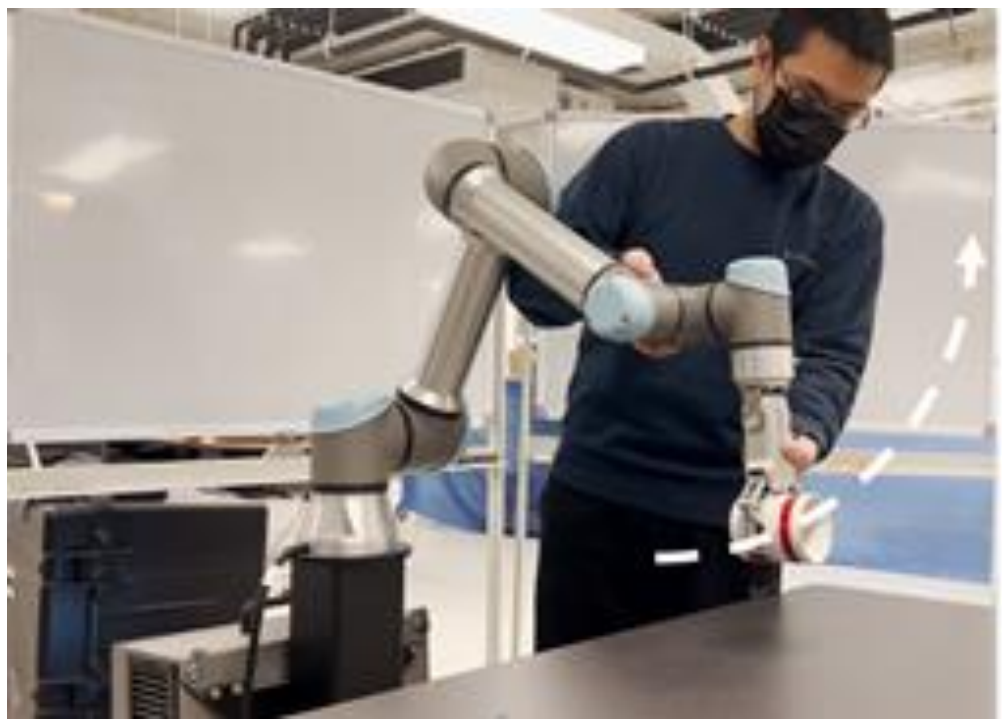

(d)

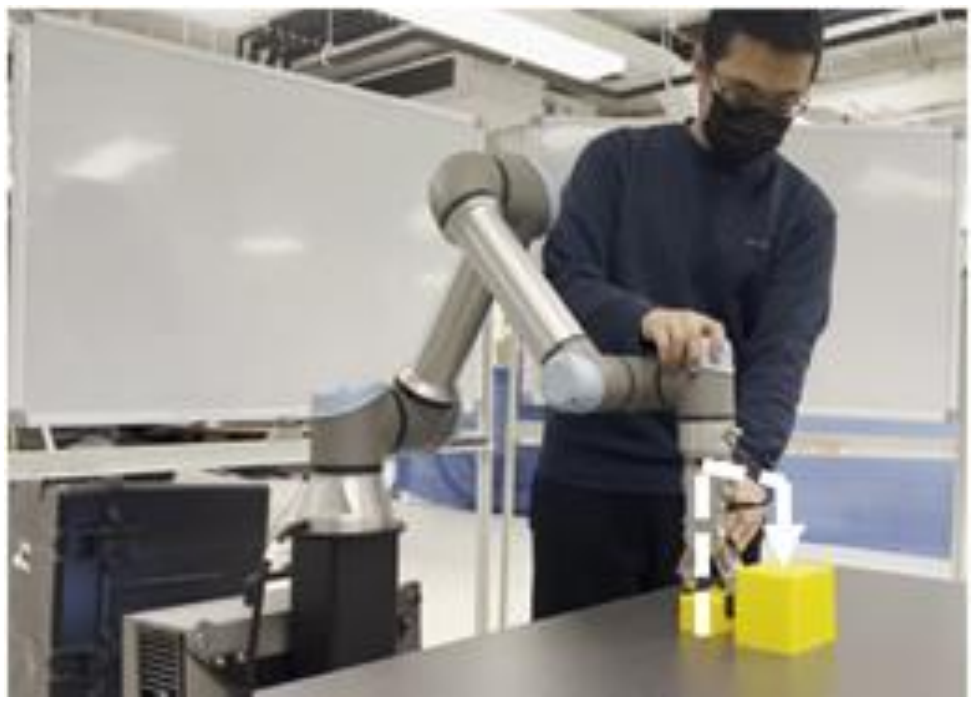

(e)

**Figure. 5** Kinesthetic demonstrations of 5 most common tasks. (a) Screwing Task 1: The end-effector is required to twisting the screw driver 90 degrees clockwise while moving straight forward to the goal position. (b) Screwing Task 2: The end-effector is required to twisting the screw driver 90 degrees anti-clockwise while moving straight forward to the goal position. (c) Pouring Task: The end-effector is required to hold the cup horizontally at first, then turn down 90 degrees. (d) Filling Task: The end-effector is required to hold the cup vertical to the ground at first, then turn up 90 degrees. (e) Stacking Task: The end-effector is required to stack the small block up the big block. The orientation of the small block should be kept upward during the stacking.

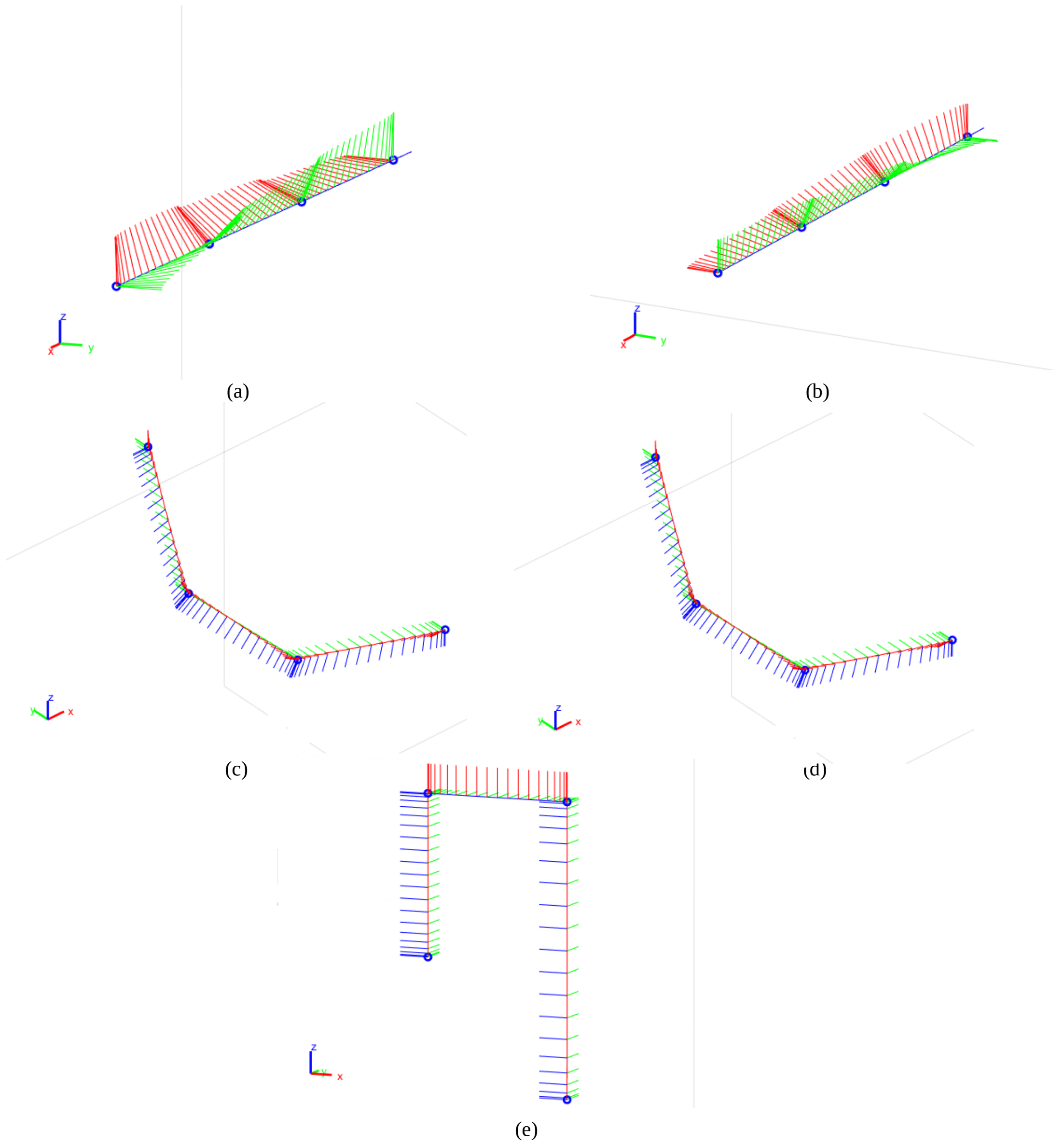

**Figure. 6** Human demonstrations in $SE(3)$. (a) Screwing task 1. (b) Screwing task 2. (c) Pouring task. (d) Filling task. (e) Stacking task.

Using forward kinematics, corresponding configurations of end-effector in $SE(3)$ are shown in Fig. 6. Using Eqn. (14), features of human demonstrations are saved as $\boldsymbol{LB} = \{\boldsymbol{HD}_1, \boldsymbol{HD}_2, \dots, \boldsymbol{HD}_5\}$ in the library shown in Table 5.

**Table 5.** Library of human demonstrated features

| | | $\boldsymbol{HD}_1$ | $\boldsymbol{HD}_2$ | $\boldsymbol{HD}_3$ | $\boldsymbol{HD}_4$ | $\boldsymbol{HD}_5$ |
|---|---|---|---|---|---|---|
| $\boldsymbol{\delta}_1$ | $\boldsymbol{d}_t^{\boldsymbol{\delta}_1}$ | (0,1,0,0) | (0,1,0,0) | (0,0.7,0,-0.7) | (0,-0.7,0,0.7) | (0,0.7,0,-0.7) |
| | $\boldsymbol{d}_r^{\boldsymbol{\delta}_1}$ | (0.7,0,0,-0.7) | (0.7,0,0,0.7) | (0.7,0,0.7,0) | (0.7,0,-0.7,0) | (1,0,0,0) |
| $\boldsymbol{\delta}_2$ | $\boldsymbol{d}_t^{\boldsymbol{\delta}_2}$ | (0,1,0,0) | (0,1,0,0) | (0,0.5,0,-0.8) | (0,-0.8,0,0.5) | (0.0.7,0,0.7) |
| | $\boldsymbol{d}_r^{\boldsymbol{\delta}_2}$ | (0.9,0,0,-0.4) | (0.8,0,0,0.5) | (0.8,0,0.5,0) | (0.8,0,-0.5,0) | (1,0,0,0) |
| $\boldsymbol{\delta}_3$ | $\boldsymbol{d}_t^{\boldsymbol{\delta}_3}$ | (0,1,0,0) | (0,1,0,0) | (0,0.2,0,-0.9) | (0,-0.9,0,0.2) | (0,0,0,1) |
| | $\boldsymbol{d}_r^{\boldsymbol{\delta}_3}$ | (1,0,0,-0.2) | (0.9,0,0,0.3) | (0.9,0,0.2,0) | (0.9,0,-0.2,0) | (1,0,0,0) |

### *6.2. Training the RL-based Motion Planner in SE(3)*

In order to obtain a general motion planning policy in $SE(3)$ for the assemble and loading/unloading scenario with the features library, we implement **Algorithm 1** to train a Q-table initialized with random Q values in Matlab using a 4-core 4.0GHz Intel Core i7 processor. Parameters for training are listed in Table 6. 20 new tasks are used during the training, each of which has four critical configurations. Positions of these critical configurations are randomly generated within a $50 \times 50 \times 50\ cm^3$ workspace. Corresponding Euler angles of each critical configuration are also randomly selected from a set $\{-\pi, -\frac{\pi}{2}, 0, \frac{\pi}{2}, \pi\}$. The total training episode for each new task is set to be 100. The total computation time is 1927.42 seconds.

To monitor the training process, accumulated rewards for each new task are recorded every two iterations. The average accumulated reward for all 20 new tasks is shown in Fig. 7. Although the training rewards are noisy before 50 episodes, the underlying trend is that the rewards are increasing with training episodes. It can be observed that the reward reaches a steady level after around 50 episodes. This indicates that a steady motion planning policy in $SE(3)$ that can map appropriate features of human demonstrations to new tasks with semantically similar features is generated for the assemble and loading/unloading scenario.

**Table 6**. Parameters for Training

| **Parameters** | $\triangle\,\alpha$ | $\triangle\,\beta$ | $\gamma$ | $\varepsilon$ |
|---|---|---|---|---|
| **Value** | 0.5 | -0.9 | 0.9 | 0.8 |

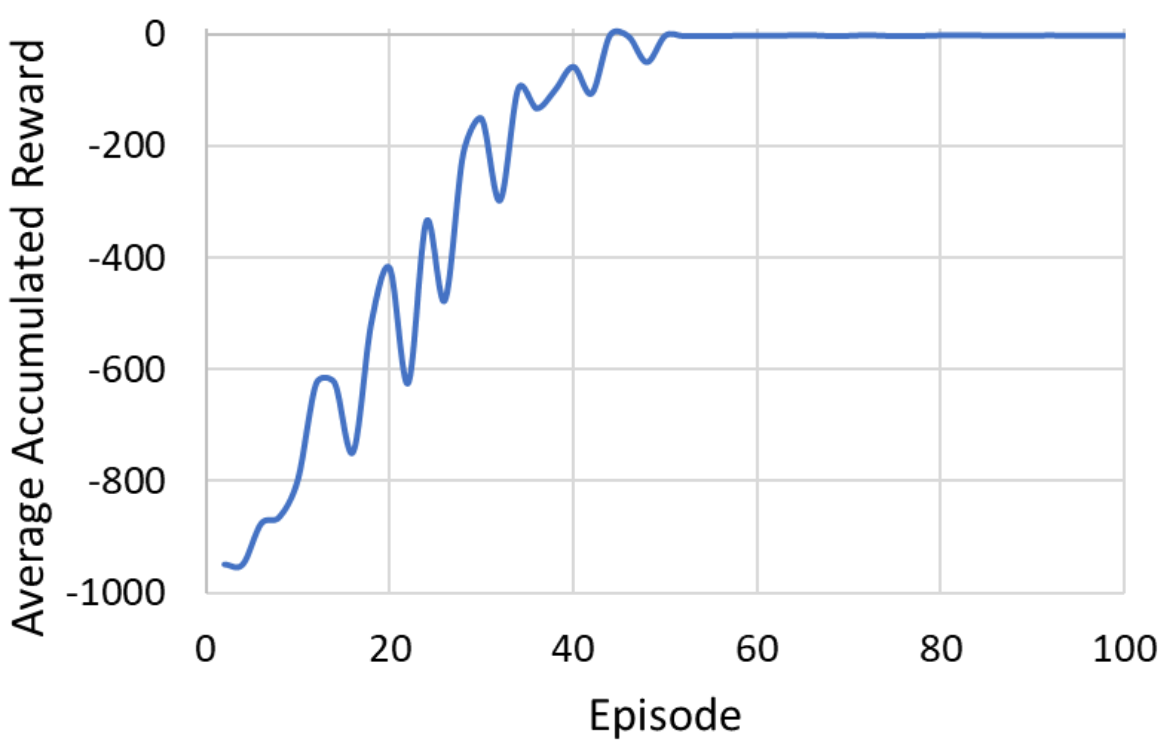

**Figure. 7** Training process for motion plans in $SE(3)$

*6.3. Evaluation of the Trained Motion Planning Policy*

To evaluate the performance of the trained motion plan policy in $SE(3)$ for the assemble and loading/unloading scenario, three new tasks, namely, a transferring task, a filling-and-pouring task, and an assembling task, are used as examples to demonstrate the method. The trained Q-table from section 6.2 is used as the input to **Algorithm 2** to generate motion plans for new unseen tasks in $SE(3)$. The inverse kinematics as discussed in Section 5 is used to calculate the final motion plan in joint space $\mathcal{J}$. For each task, 20 experiment trials are conducted to evaluate the successful executions.

**Transferring Task**: In this task, the end-effector is required to transfer a cup of water while avoiding an obstacle shown as red cube in Fig. 8. The dimension of the obstacle is $20 \times 20 \times 20\ cm^3$ and the position of its center is $(-0.3, -0.1, 0.2)$. To avoid this obstacle, a safety protocol is assumed given as a $40 \times 40 \times 40\ cm^3$ safety shell (shown as transparent purple cube in Fig. 8 with the same center position as the cubic obstacle) that the manipulator cannot penetrate through. Knowing these environment constraints and the task requirement on moving the cup of water from the starting position to the goal position, a sample of user defined critical configurations, $con_1, con_2, con_3, con_4$ are summarized in Table 7, where $0.1 \leq z \leq 0.5$. In these critical configurations, $con_1$ and $con_4$ describe the end-effector starting and the ending positions and orientations, $con_2$ and $con_3$ are two intermediate critical configurations selected on edges of the safety shell. Note that different users may have different task specifications depending on their understandings of the task and environment constraints.

The result shows that all 20 trials are successfully executed, where each of the 20 feasible motion plans (shown as the path composed of the small triads) complies with the task requirement that the end-effector maintains the same orientation of $(0, -\pi/2, 0)$ to prevent the spill out. One of the 20 motion plans are demonstrated as the yellow line in Fig. 8, where the orientations of the small triads along each path represent the orientations of the end-effector. Execution of the corresponding motion plan in the joint space is shown in Fig. 8 (b). It is noticed that the feature of the human demonstrated stacking task is learned and mapped for this transferring task. In this experiment, both explicit task constraints (position and orientation constraints of critical configurations) and implicit task constraints (keeping the orientation of the end-effector) are satisfied. This experiment demonstration the effectiveness of the motion plan policy trained in section 6.2, which can be used as motion plans to avoid the obstacles in $SE(3)$ if users have sufficient knowledge about the task and the environment and can properly infuse the knowledge in task specification based on the syntax we defined.

**Table 7**. Critical configurations of Task 3

| | $con_1$ | $con_2$ | $con_3$ | $con_4$ |
|---|---|---|---|---|
| $\boldsymbol{P}$ | (-0.5,0,0.3) | (-0.4,0.2,$z$) | (0,0.2,$z$) | (0.1,0,0.3) |
| $\boldsymbol{\theta}$ | $(0, -\pi/2, 0)$ | $(0, -\pi/2, 0)$ | $(0, -\pi/2, 0)$ | $(0, -\pi/2, 0)$ |

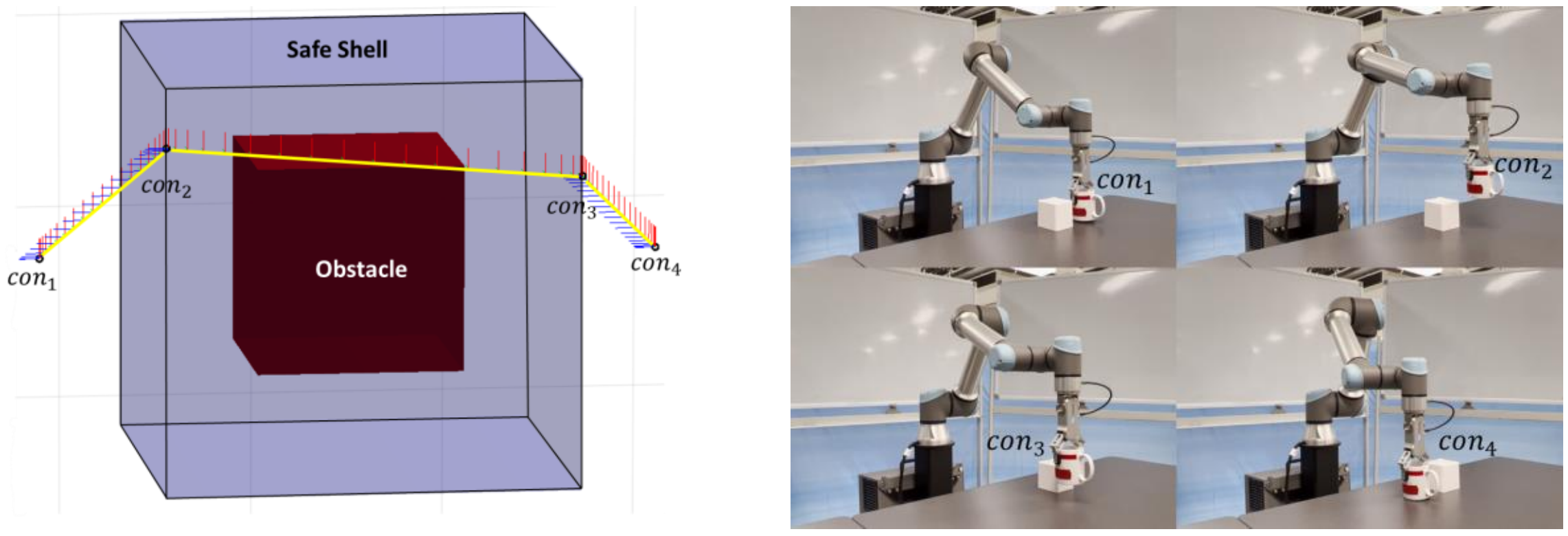

(a) (b)

**Figure. 8** Motion plans for the transferring task. (a) The motion plan in $SE(3)$. (b) Execution the motion plan in $\mathcal{J}$.

**Filling-and-Pouring Task**: As shown in Fig. 9 (a), in this task, the end-effector is required to fill water to Cup 1, then go through another two critical positions, and reach a goal position above Cup 2, and finally pour water to Cup 2. The location of Cup 2 is on the surface of a desk within a workspace of $20 \times 20\ cm^2$. We can specify 5 critical configurations based on the described task. Sample critical configurations from $con_1$ to $con_5$ are presented in Table 8, where $-\pi \leq \gamma \leq \pi$, $-0.5 \leq x \leq 0.7$, $-0.2 \leq y \leq 0$. It is noted that, in this task, only the orientation of the initial configuration and goal configuration are specified with specific Euler angles. When moving the cup from $con_2$ to $con_4$, the end-effector is only required upward without any specific constraints in the yaw angle.

**Table 8**. Critical configurations of $\boldsymbol{TK}$

| | $con_1$ | $con_2$ | $con_3$ | $con_4$ | $con_5$ |
|---|---|---|---|---|---|
| $\boldsymbol{P}$ | (-0.4,-0.1,0) | (-0.5,0.1,0.1) | (-0.7,-0.1,0.1) | $(x, y, 0.1)$ | $(x, y, 0)$ |
| $\boldsymbol{\theta}$ | $(0,-\pi,0)$ | $(0, -\pi/2,\gamma)$ | $(0, -\pi/2,\gamma)$ | $(0, -\pi/2,\gamma)$ | $(\pi/2,0, -\pi/2)$ |

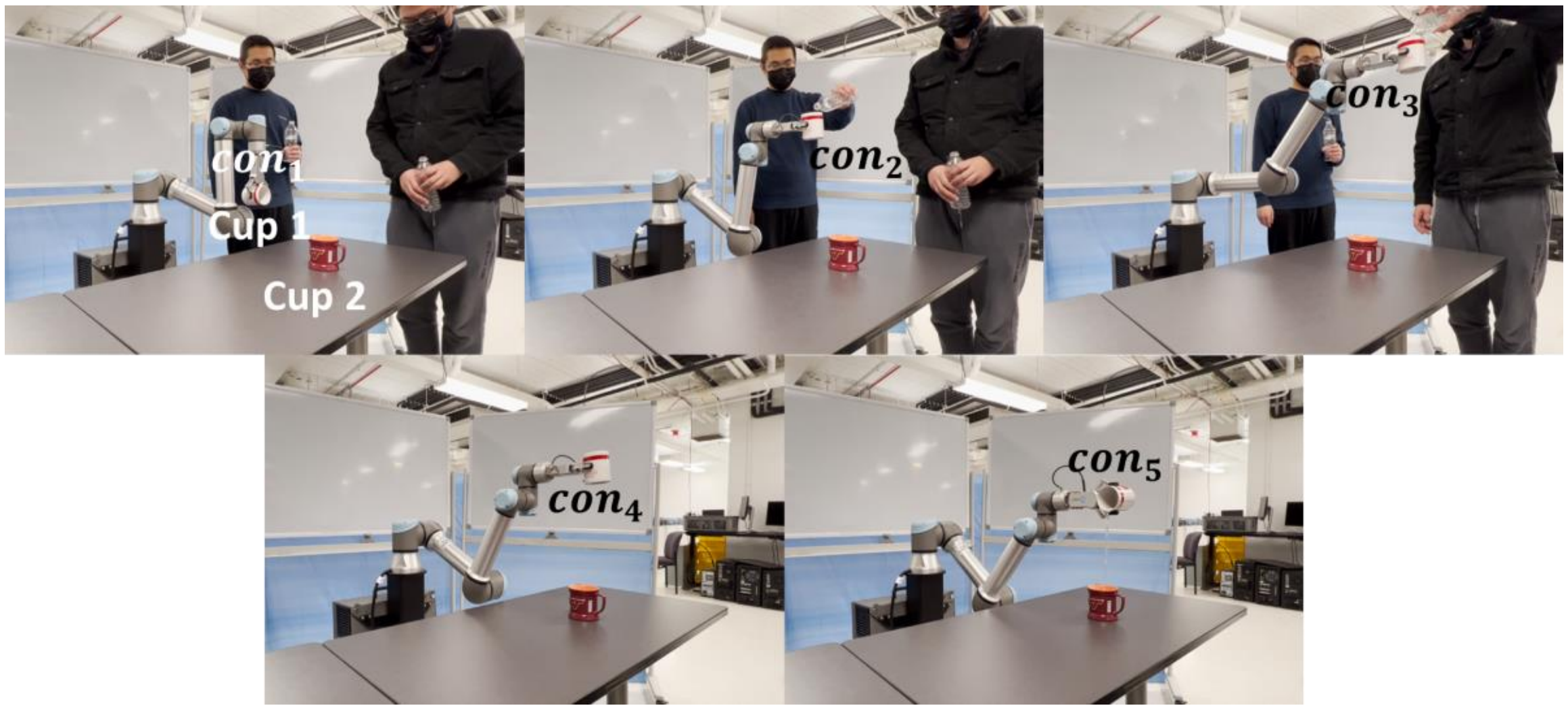

(a)

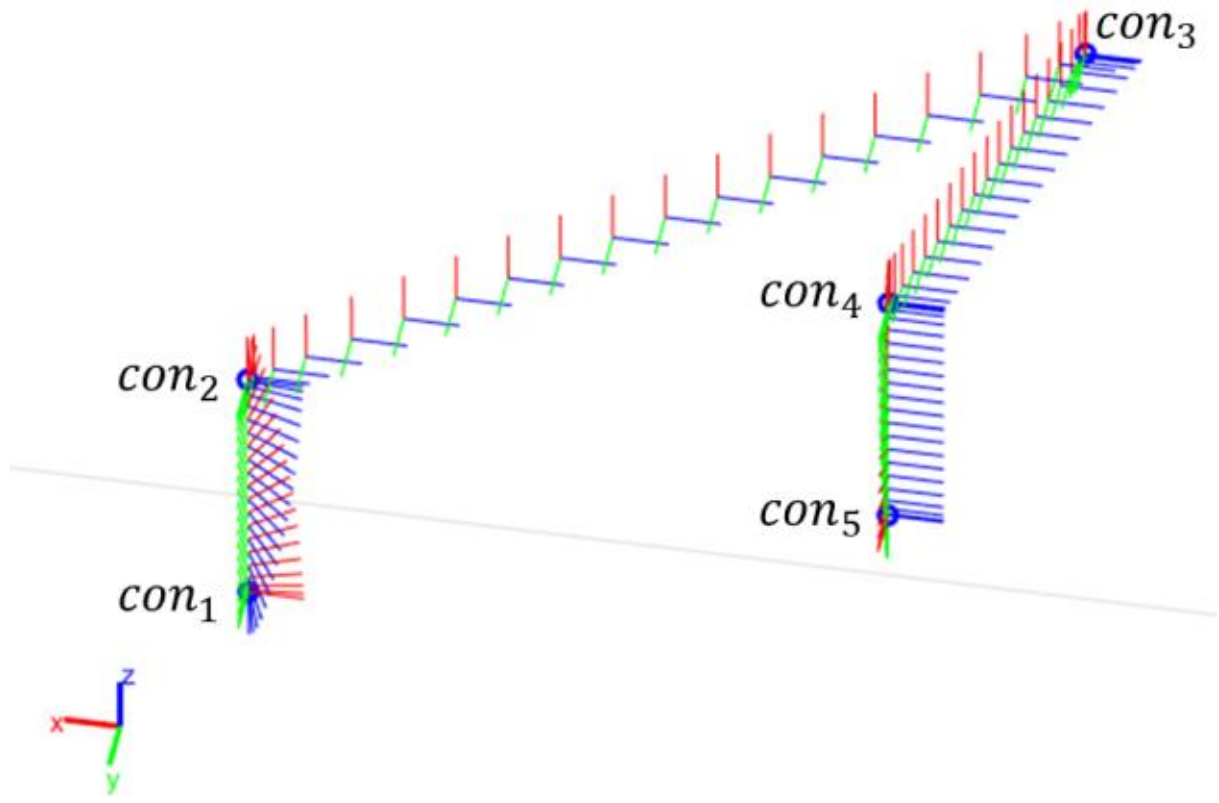

(b)

**Figure. 9** Generate motion plans for the filling-and-pouring task. (a) The motion plan in $SE(3)$ (b) The final execution of the motion plan in $\mathcal{J}$.

For 20 experiment trials with various locations of Cup 2, all experiments are successfully performed in $\mathcal{J}$. We use the motion plan in $\mathcal{J}$ and $SE(3)$ for one trial as an example as shown in Fig. 9 (a) and (b) to illustrate the result. It is noticed that the feature of 3 human demonstrated tasks, namely filling, stacking, and twisting, are learned and mapped to the segment between $con_1$ and $con_2$, the segment between $con_2$ and $con_4$, and the segment between $con_4$ and $con_5$, respectively. In this experiment, the proposed method can identify and compose the appropriate features in the human demonstration library to perform a new task.

**Assembling Task**: In this task, the end-effector needs to disassemble a screw from Assembly Hole 1, then place the screw into Assembly Hole 2, and finally fasten the screw. The location of Assembly Hole 1 and Assembly Hole 2 are $(0, 0.5, 0.6)$ and $(0.5, 0, 0.6)$ as shown in Fig. 10 (a). To transfer the screw from Assembly Hole 1 to Assembly Hole 2, the end-effector is required to hold the screw horizontally and turn 90 degrees anti-clockwise. Based on the task requirement, the critical configurations can be specified in Table 9.

**Table 9.** Critical configurations of the assembling task

| | $con_1$ | $con_2$ | $con_3$ | $con_4$ |
|---|---|---|---|---|
| $\boldsymbol{P}$ | (0,0.5,0.6) | (0.1,0.5,0.6) | (0.5,0.1,0.6) | (0.5,0,0.6) |
| $\boldsymbol{\theta}$ | $(-\pi/2,0,\pi/2)$ | $(0,-\pi/2,0)$ | $(0,-\pi/2,\pi/2)$ | $(\pi,0,-\pi/2)$ |

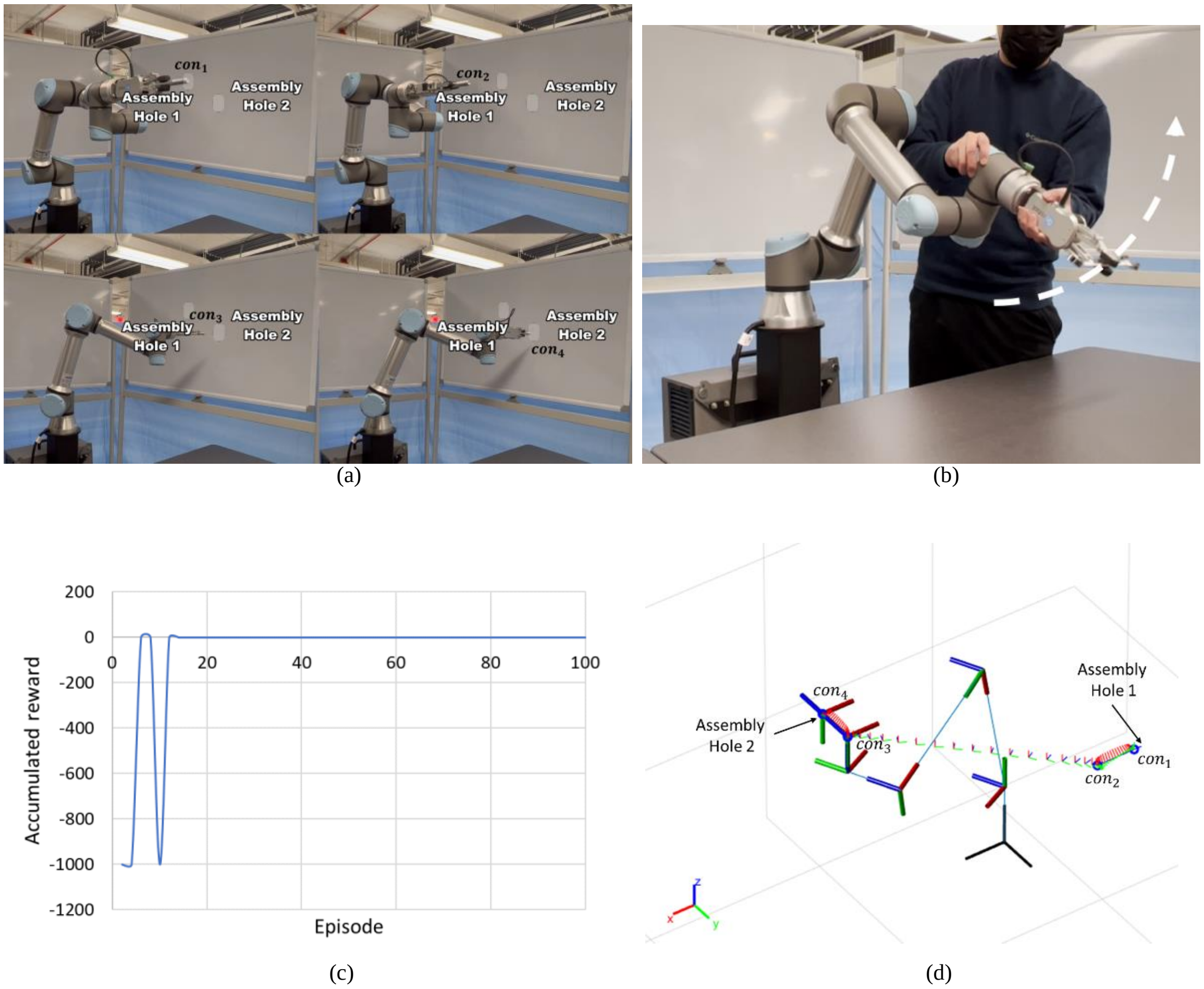

**Figure. 10** Generate motion plans for the assembling task after the feature of a new human demonstration is added to the library. (a) Critical configurations and the motion plan in $\mathcal{J}$ for the assembling task after an additional human demonstration is provided. (b) Additional human demonstration for picking up a span. (c) The training process of the assembling task after the feature of a new human demonstration is added. (d) Motion plan in $SE(3)$.

By applying the same trained general motion plan policy in section 6.2, no successful motion plan can be generated, which indicates additional human demonstrations are needed. A closer examination reveals that none of the five features saved in the library is semantically similar to the feature of the task segment between $con_2$ and $con_3$, which requires a 90-degree rotation about its body-fixed x-axis clockwise. Therefore, additional human demonstration is requested for this feature.

With this additional human demonstration shown in Fig. 10 (b) added in the library, the motion planning policy is retrained using **Algorithm 1** with the Q-table trained in section 6.2. As shown in Fig. 10 (c), the accumulated reward reaches a steady value after around 10 iterations. Then by applying the newly trained policy, the motion plan in $SE(3)$ is generated as shown in Fig. 10 (d). The result shows that the features of the human demonstrated twisting task 1 and task 2 are learned and mapped to the task segment between $con_1$ and $con_2$, and the segment between $con_3$ and $con_4$, respectively. The feature of the newly added human demonstration is learned and mapped to the task segment between $con_2$ and $con_3$. Corresponding motion plan in $\mathcal{J}$ is shown in Fig. 10 (a).

To summarize, the case study results demonstrate the effectiveness of the proposed RL-based user-guided motion planning method in learning and mapping appropriate features of human demonstrations to new tasks and generating motion plans in the joint space for semantically similar tasks. The proposed method can also request additional human demonstrations when new task features cannot be found in the human demonstration library.

## 7. Conclusion and Future Work

In this paper, we present a novel method for robot learning from human demonstrations based on RL-based motion planning. A task specification scheme is first developed for users to provide necessary kinematic information about task and environment constraints. A human demonstration library for specific working scenarios is built through recording and storing the common actions by utilizing the existing recording capability for modern robots. By abstracting features from human demonstrations and tasks, the task-space RL-based motion planner can effectively identify, learn, and compose the appropriate demonstrated features to perform new tasks that comply with the task requirements and environment constraints. Followed by inverse kinematics, motion plans in joint space can be obtained

In future work, we plan to explore using different hardware architecture since the motion planning is done in $SE(3)$. Joint limit consideration will be further studied. In addition, we will extend this work to integrate with the human motion perception, recognition, and prediction for realistic implementation.

## Acknowledgment

This work was supported by the U.S. National Science Foundation (NSF) Grant. CMMI1853454. We also appreciate Cheng Zhu's effort in experiments.